%% file: cat_paper.tex
\documentclass[useAMS,usenatbib]{mn2e}
\usepackage{graphicx,longtable,amssymb,topcapt,txfonts,rotating,lscape}

\begin{document}

\title[Cores in IRDCs from l = 300 to 330\,\degr]{Cores in
Infra-Red Dark Clouds (IRDCs) seen in the Hi-GAL
survey between l = 300$\,\degr$ and l = 330$\,\degr$}
\author[Wilcock, L.~A. et al.]{L.~A. Wilcock$^{1}$, D. Ward-Thompson$^{1}$, 
J.~M. Kirk$^{1}$, D. Stamatellos$^{1}$, A. Whitworth$^{1}$, \newauthor D. Elia$^{2}$, G.~A. 
Fuller$^{3}$, A. DiGiorgio$^{2}$, M. J. Griffin$^{1}$, S. Molinari$^{2}$, 
P. Martin$^{4}$, J.~C. Mottram$^{5}$, \newauthor N. Peretto$^{3,6}$, M. Pestalozzi$^{2}$, 
E. Schisano$^{2}$, R. Plume$^{7}$, H.~A. Smith$^{8}$ \& M.~A. Thompson$^{9}$ \\ 
$^{1}$School of Physics and Astronomy, Cardiff University, Queen's 
Buildings, Cardiff, CF24 3AA, UK \\
$^{5}$ School of Physics, University of Exeter, Stocker Road, Exeter, 
EX4 4QL, UK \\
$^{3}$ Jodrell Bank Centre for Astrophysics, School of Physics and 
Astronomy, University of Manchester, Manchester, M13 9PL, UK \\
$^{2}$ Instituto di Fisica dello Spazio Interplanetario, CNR, via Fosso del 
Cavaliere, I-00133 Roma, Italy \\
$^{4}$ Canadian Institute for Theoretical Astrophysics, University of 
Toronto, Toronto, Canada, M5S 3H8 \\
$^{6}$  Laboratoire AIM, CEA/DSM-CNRS-Universit\'e Paris Diderot, 
IFRU/Service d'Astrophysique, C.E. Saclay, 
Orme des merisiers, \\ 91191 Gif-sur-Yvette, France \\
$^{7}$ University of Calgary, Dept Physics-Astronomy, Calgary, AB 
T2N 1N4, Canada \\
$^{8}$ Harvard-Smithsonian Center for Astrophysics, 60 Garden Street, 
Cambridge, MA, 02138, USA \\
$^{9}$ Centre for Astrophysics Research, Science and Technology 
Research Institute, University of Hertfordshire, AL10 9AB, UK \\}
\maketitle

\label{firstpage}

\begin{abstract}
We have used data taken as part of the \textit{Herschel} infrared
Galactic Plane survey (Hi-GAL) to study 3171 infrared-dark cloud
(IRDC) candidates that were identified in the mid-infrared
(8~$\mu$m) by \textit{Spitzer} (we refer to these as `\textit{Spitzer}-dark' regions).
They all lie in the range $l$$=$$300-330$\,$\degr$ and $|b|\le1$\,\degr. 
Of these, only 1205 were seen in 
emission in the far-infrared (250--500~$\mu$m) by \textit{Herschel} 
(we call these `\textit{Herschel}-bright' clouds).
It is predicted that a dense cloud will not only be
seen in absorption in the mid-infrared, but will also
be seen in emission in the far-infrared at the longest \textit{Herschel} 
wavebands (250--500~$\mu$m).
If a region is dark at all wavelengths throughout the mid-infrared
and far-infrared, then it is most likely to be simply a region of
lower background infrared emission (a `hole in the sky').
Hence, it appears that previous surveys, based on \textit{Spitzer} and 
other mid-infrared data alone, may have over-estimated the total IRDC
population by a factor $\sim$2.
This has implications for estimates of the star formation rate in IRDCs in the Galaxy.
We studied the 1205 \textit{Herschel}-bright IRDCs at 250~$\mu$m, and found that
972 of them had at least one clearly defined 250-$\mu$m peak, 
indicating that they contained one or more dense cores. Of these, 
653 (67\,per\,cent) contained an 8-$\mu$m point source somewhere within the cloud,
149 (15\,per\,cent) contained a 24-$\mu$m point source but no 8-$\mu$m source, and 
170 (18\,per\,cent) contained no 24-$\mu$m or 8-$\mu$m point sources. We use
these statistics to make inferences about the lifetimes of the various
evolutionary stages of IRDCs.
\end{abstract}


\section{Introduction}

Infrared dark clouds
(IRDCs) were initially discovered by the \textit{MSX} \citep{carey98, egan98} 
and \textit{ISO} \citep{perault96}
surveys as dark regions seen against the mid-infrared (MIR) background. The 
densest IRDCs may eventually form massive stars (e.g. \citealt{perault96,kauffmann10}),
and are presumed to represent the earliest observable stage of 
high mass star formation. Some IRDCs contain 
cores without embedded protostars,
which are believed to be the high mass equivalent of 
low-mass prestellar cores \citep{rathborne06,chambers09}. 

\begin{figure*}
\begin{center}
\includegraphics[width=85mm]{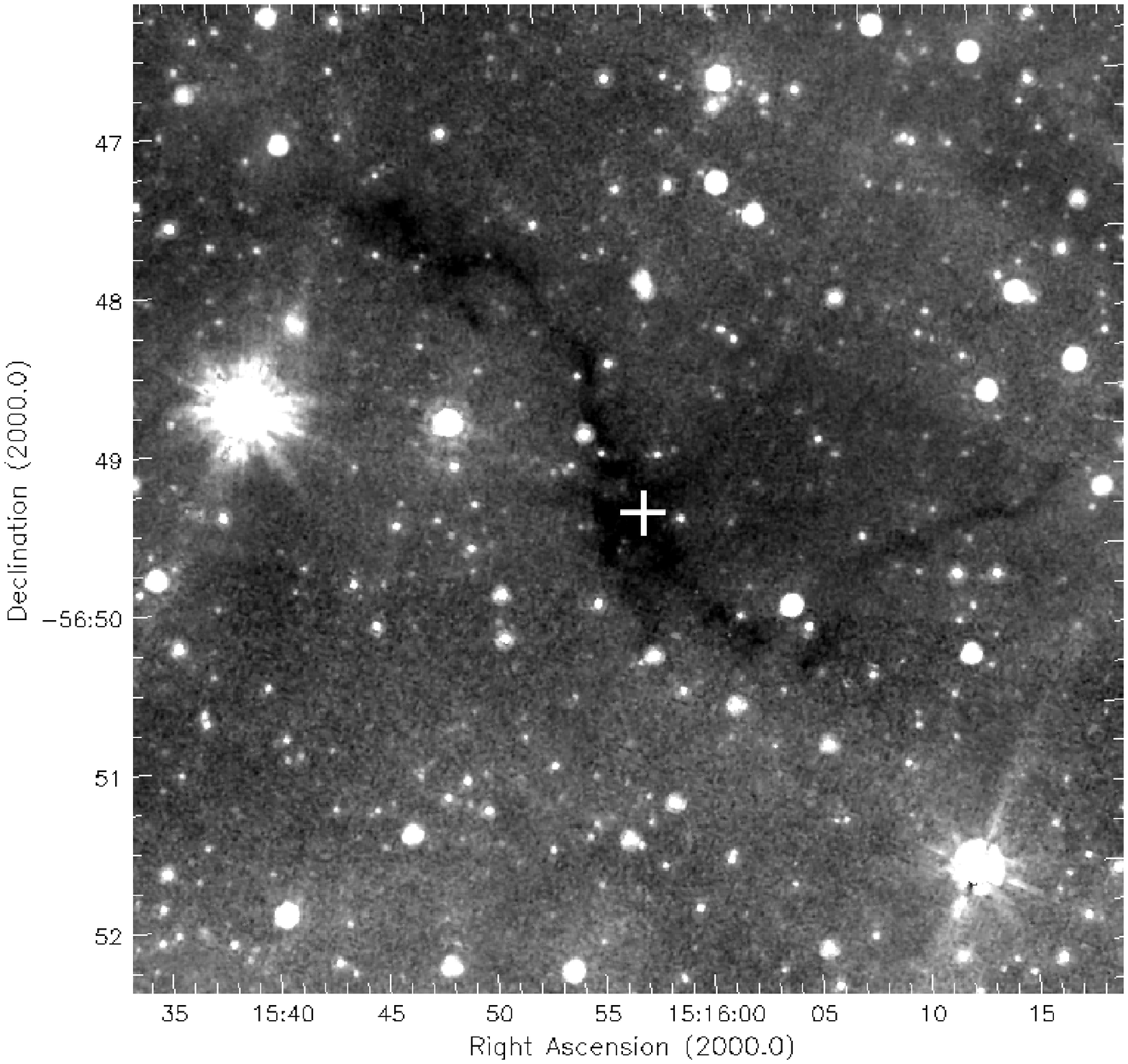}
\includegraphics[width=85mm]{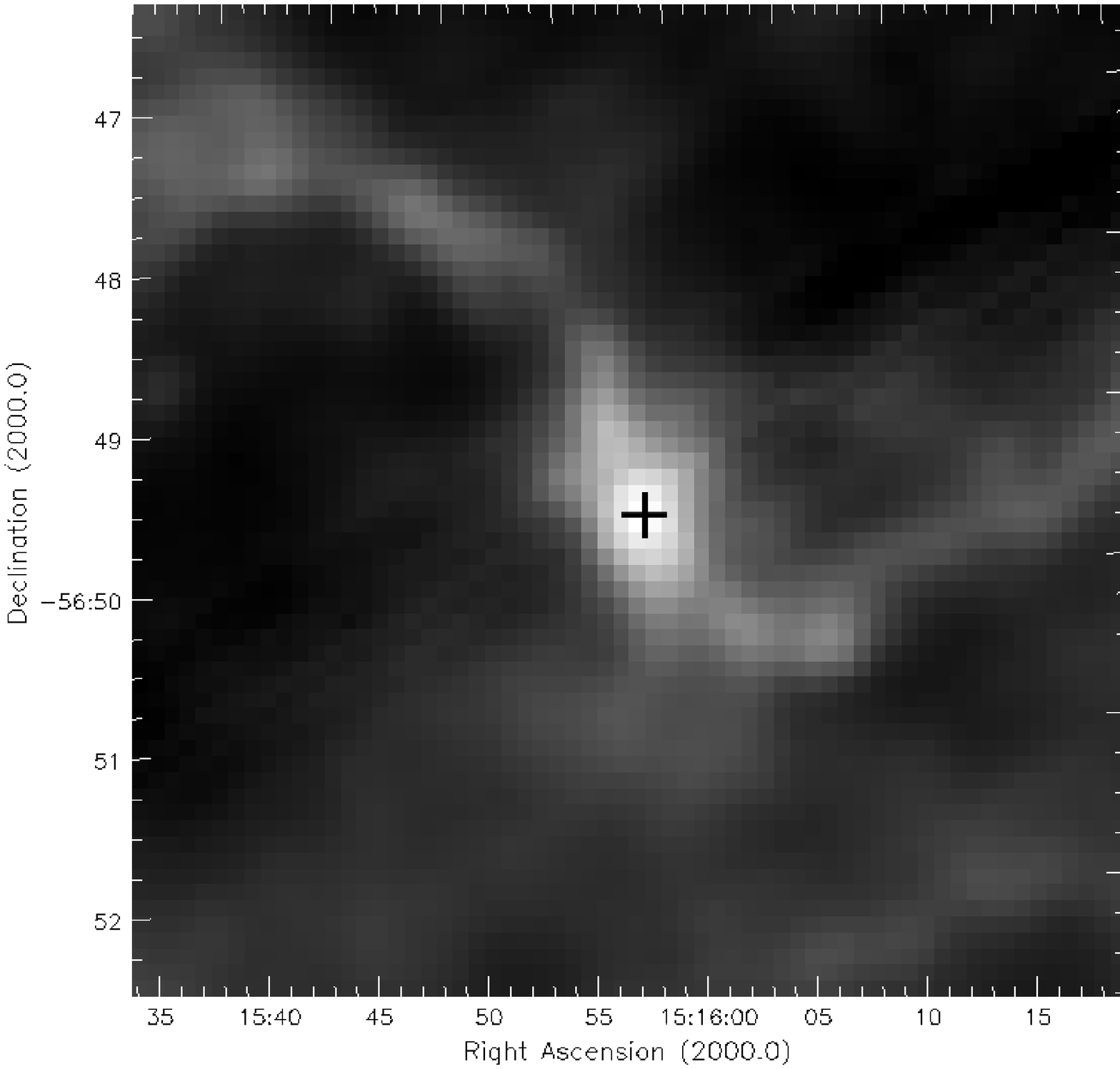}
\end{center}
\caption{G321.753+0.669. Left panel: 8~$\mu$m. Right panel: 250~$\mu$m.
The position of the candidate IRDC is shown with a cross. This is
an example of a \textit{Spitzer}-dark and \textit{Herschel}-bright cloud. 
Note how the same structure that is seen in absorption (black) in
the left panel is seen in emission (white) in the right panel.
This is believed to be a genuine IRDC.} \label{realirdc}
\end{figure*}

Observations of IRDCs and their cores have shown them to have low 
temperatures (T$<$25\,K; e.g. \citealt{egan98, teyssier02}) 
and high densities
(n$_H$$>$$10^5 $\,cm$^{-3}$; \citealt{egan98,carey98,carey00}). 
Kinematic calculations have shown that IRDCs are typically at
a galactocentric distance of 6\,kpc in the fourth quadrant and 5\,kpc in the first 
quadrant of the Galaxy, matching the location of the Scutum-Centaurus arm
\citep{jackson08}. IRDCs have masses between a few hundred and several 
thousand solar masses, while masses of cores within IRDCs tend to
lie in the range 10--1000\,M$_{\odot}$ \citep{rathborne06,swift09,peretto10b,devine11,
wilcock11,zhang11}.

There are currently two published catalogues of candidate IRDCs, namely 
\citet{simon06c} and \citet{peretto09} -- hereafter PF09.
The former used 8.3-$\mu$m 
\textit{MSX} data to identify 10931 candidate
IRDCs, while the latter used \textit{Spitzer} 8-$\mu$m data to
find 11303 candidate IRDCs.

\citet{chambers09} use a selection of the \citet{simon06c} IRDCs to 
propose a hypothetical evolutionary sequence wherein
cores evolve from a quiescent to an active
phase and finally into a red core. 
The quiescent cores contain no 
MIR activity and are likely to be starless, not
yet undergoing any star formation. As the core evolves it enters the 
active phase and begins to show tracers indicating
star formation. These tracers include 24-$\mu$m emission, maser 
emission and extended green objects (EGOs; \citealt{cyganowski08}), which are regions of 
enhanced 4.5-$\mu$m emission thought to be shocked H$_2$ gas
\citep{debuizer10} and 
thus charactistic of an outflow, sometimes called `green 
fuzzies' \citep{chambers09}. 

A core is said to be in the final, red, stage when it 
shows PAH emission and is therefore bright at 8~$\mu$m. PAH 
emission is seen in regions with
high ultra-violet radiation fields. Hence, red cores may 
contain hyper-compact or ultra-compact HII regions. Other studies have also 
used these, or similar, star formation tracers when studying IRDCs 
(e.g. \citealt{jimenez10,battersby11,devine11}).

\citet{chambers09} had a sample of 190 candidate IRDCs, and they found
$\sim$54\,per\,cent to be quiescent,
$\sim$25\,per\,cent to be active and 
$\sim$21\,per\,cent to be red cores.
There is some debate over the lifetimes of the different stages of 
cores in IRDCs. \citet{chambers09} use the accretion timescale of
high mass star formation and find statistical lifetimes for the 
quiescent and active phases to be 3.7 and 2.0 $\times$ 10$^5$ years 
respectively. 

\citet{parsons09} had a sample of 69 \citet{simon06c} IRDCs, and
found the ratio to be $\sim$30\,per\,cent quiescent and 
$\sim$70\,per\,cent to have some form of embedded source
(either active or red in the Chambers nomenclature).
They use a different estimated 
lifetime for the embedded YSO phase and find a timescale of 
about an order of magnitude less than \citet{chambers09}
for the starless, quiescent phase.
These lifetime estimates make the 
assumption that all starless IRDC cores will eventually begin 
forming high mass stars \citep{battersby10}. They
should be viewed as no more than
order of magnitude estimates at best. 

In this paper, we use data from the \textit{Herschel} infrared 
Galactic Plane Survey (Hi-GAL; \citealt{higalb,higala}) to observe the 
IRDCs of PF09 at far-infrared (FIR)
wavelengths. 

\begin{figure*}
\begin{center}
\includegraphics[width=75mm,angle=-90]{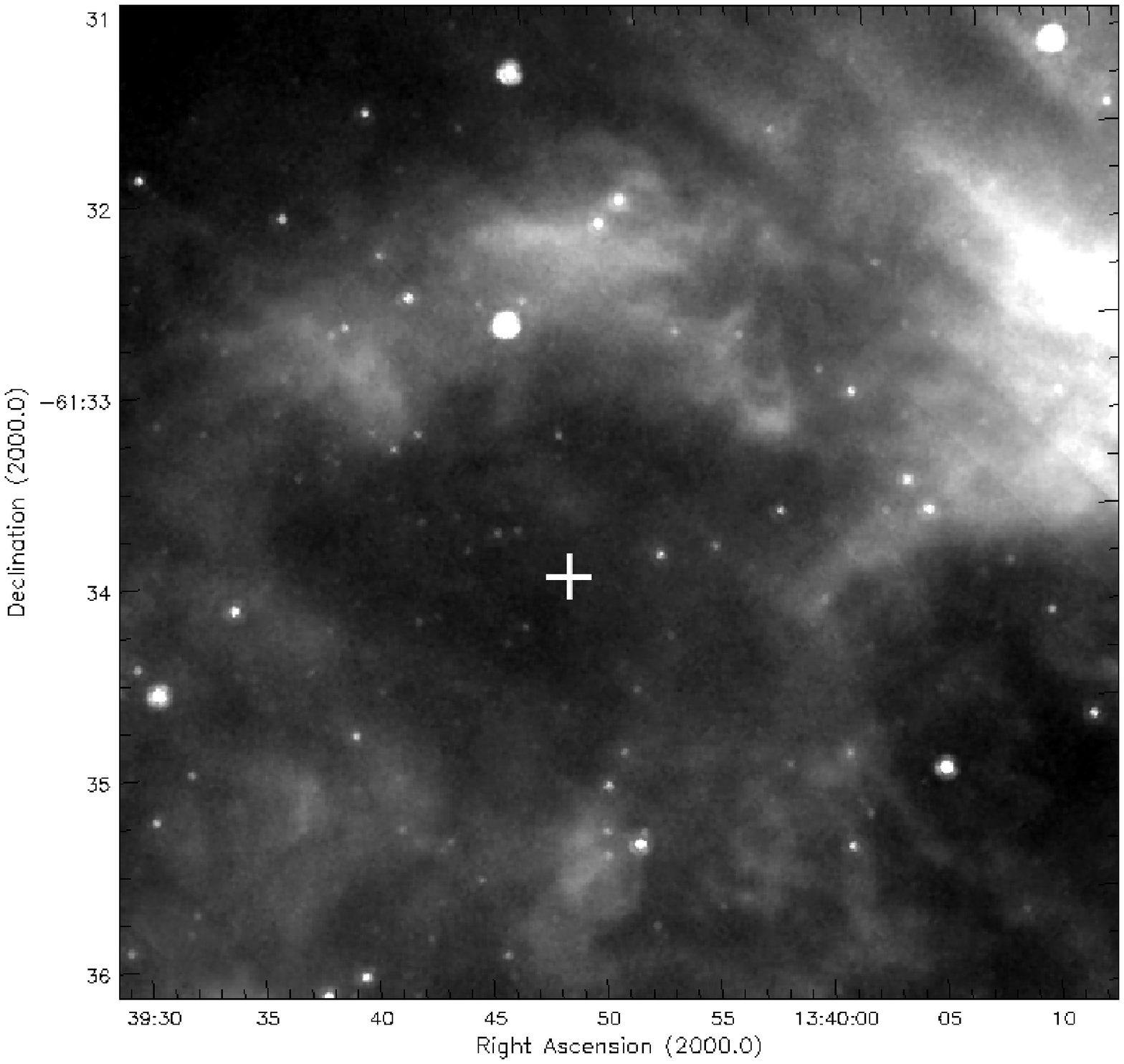}
\includegraphics[width=75mm,angle=-90]{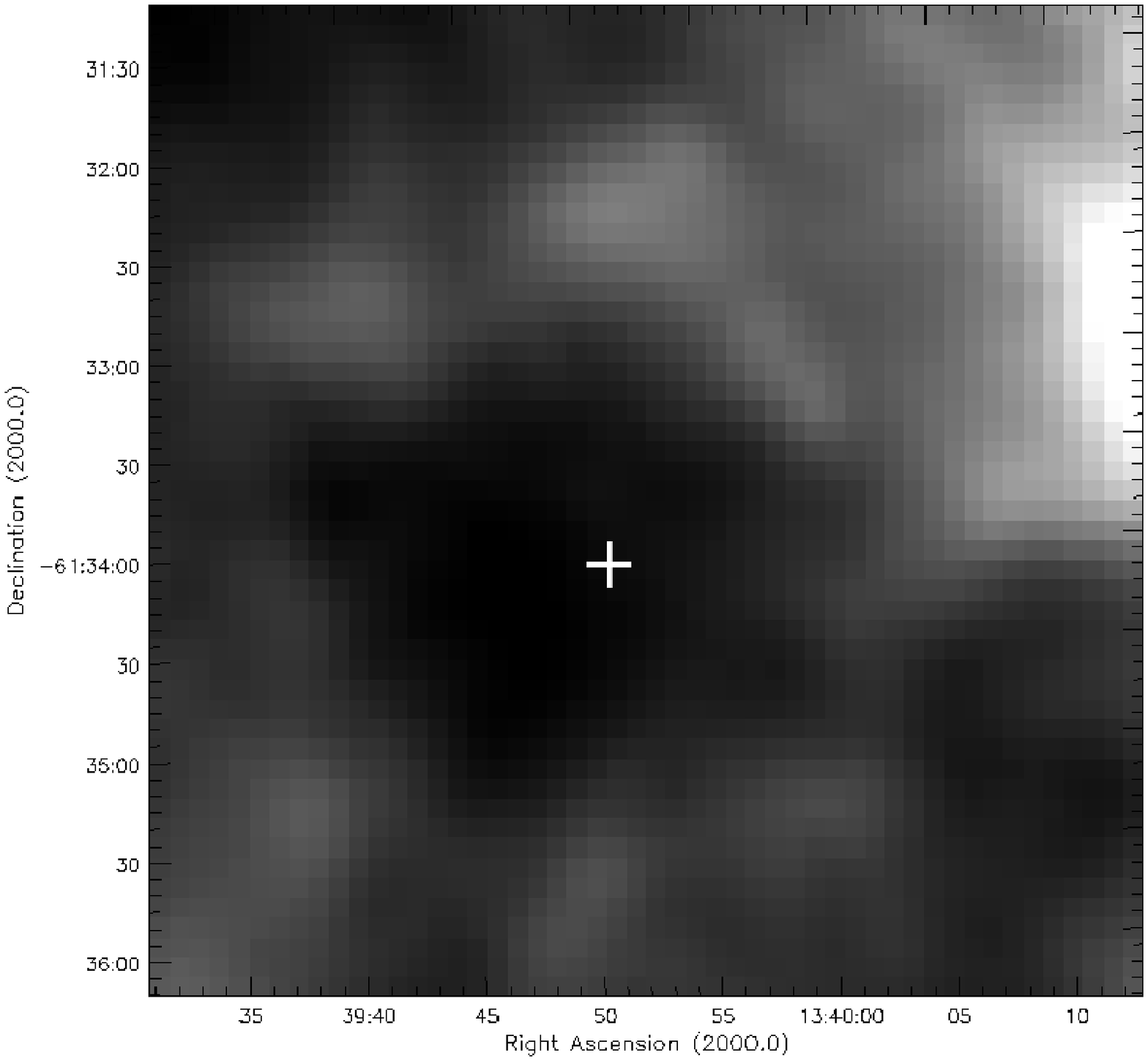}
\end{center}
\caption{G308.656+0.760. Left panel: 8~$\mu$m. Right panel: 250~$\mu$m.
The position of the candidate IRDC is shown with a cross.
This is an example of a candidate IRDC
that is \textit{Spitzer}-dark but is not \textit{Herschel}-bright. 
The same structure that appears dark (black) in the left panel is also
dark (black) in the right panel. We refer to such a candidate as \textit{Herschel}-dark. 
This is not believed to be a genuine IRDC.} \label{hole}
\end{figure*}

\section{OBSERVATIONS}\label{data}

\subsection{Herschel}

The \textit{Herschel} Space Observatory\footnote{\textit{Herschel} 
is an ESA space observatory with science instruments provided by 
European-led Principal Investigator consortia and 
with participation from NASA.}  \citep{herschel} was 
launched in May 2009, and carries three instruments: the Spectral and 
Photometric Imaging Receiver (SPIRE, \citealt{spire}); the 
Photodetector Array Camera and Spectrometer (PACS; \citealt{pacs}); 
and the Heterodyne Instrument for the Far Infrared (HIFI; 
\citealt{hifi}). It is capable of observing in the FIR 
between 55 and 671\,$\mu$m. 
The data used in this paper were taken as part of the \textit{Herschel} 
infrared Galactic Plane Survey (Hi-GAL), an Open Time Key
Project of the \textit{Herschel} Space Observatory \citep{higalb,higala}. 
Hi-GAL aims to perform a survey of the Galactic Plane 
using the PACS and SPIRE instruments. The two are used in parallel mode 
to map the Milky Way Galaxy simultaneously at five 
wavelengths (70, 160, 250, 350 and 500\,$\mu$m), with resolutions up to 
5\arcsec{} at 70\,$\mu$m.

PACS data reduction at 70 and 160~$\mu$m was 
performed using the \textit{Herschel} Interactive 
Pipeline Environment (HIPE; \citealt{hipe}), with some
additions described by \citet{pacs}. The standard deglitching 
and crosstalk correction were not used and custom procedures 
were written for drift removal \citep{traficante11}. 
SPIRE data processing at 250, 350 and 500~$\mu$m
used the standard processing methods \citep{spire}, 
with both standard deglitching and drift removal. 
In all cases, the ROMAGAL Generalised Least Squares algorithm 
\citep{traficante11} was used to produce the final maps. A more 
detailed discussion of the data reduction process is given by 
\citet{traficante11}.

\subsection{Spitzer}

The \textit{Spitzer} Space  Telescope\footnote{\textit{Spitzer} 
was operated by the Jet Propulsion 
Laboratory at the California Institute of Technology under 
a contract with NASA.} \citep{spitzer} was launched 
in August 2003, and carried three instruments: 
the Multiband Infrared Photometer for \textit{Spitzer} (MIPS; \citealt{mips}); 
the Infrared Array Camera (IRAC; \citealt{irac}); 
and the Infrared Spectograph (IRS; \citealt{irs}). 
These instruments are capable of
observing in a number of wavebands ranging between 3.6 and 160~$\mu$m.
As part of its legacy science programme, \textit{Spitzer} performed two 
surveys of the Galactic Plane. These were the MIPS
Galactic Plane Survey (MIPSGAL; \citealt{mipsgal}) and the Galactic 
Legacy Infrared Mid-Plane Survey Extraordinaire (GLIMPSE; \citealt{glimpse}). 
Here, we use the mosaics made available by the \textit{Spitzer} Science 
Centre to create 8- and 24-$\mu$m maps of the region encompassing
$300\,\degr$$\le$$l$$\le$$330\,\degr$, and $|b|$$\le$$1\,\degr$.

\begin{figure*}
\begin{center}
\includegraphics[angle=0,width=85mm]{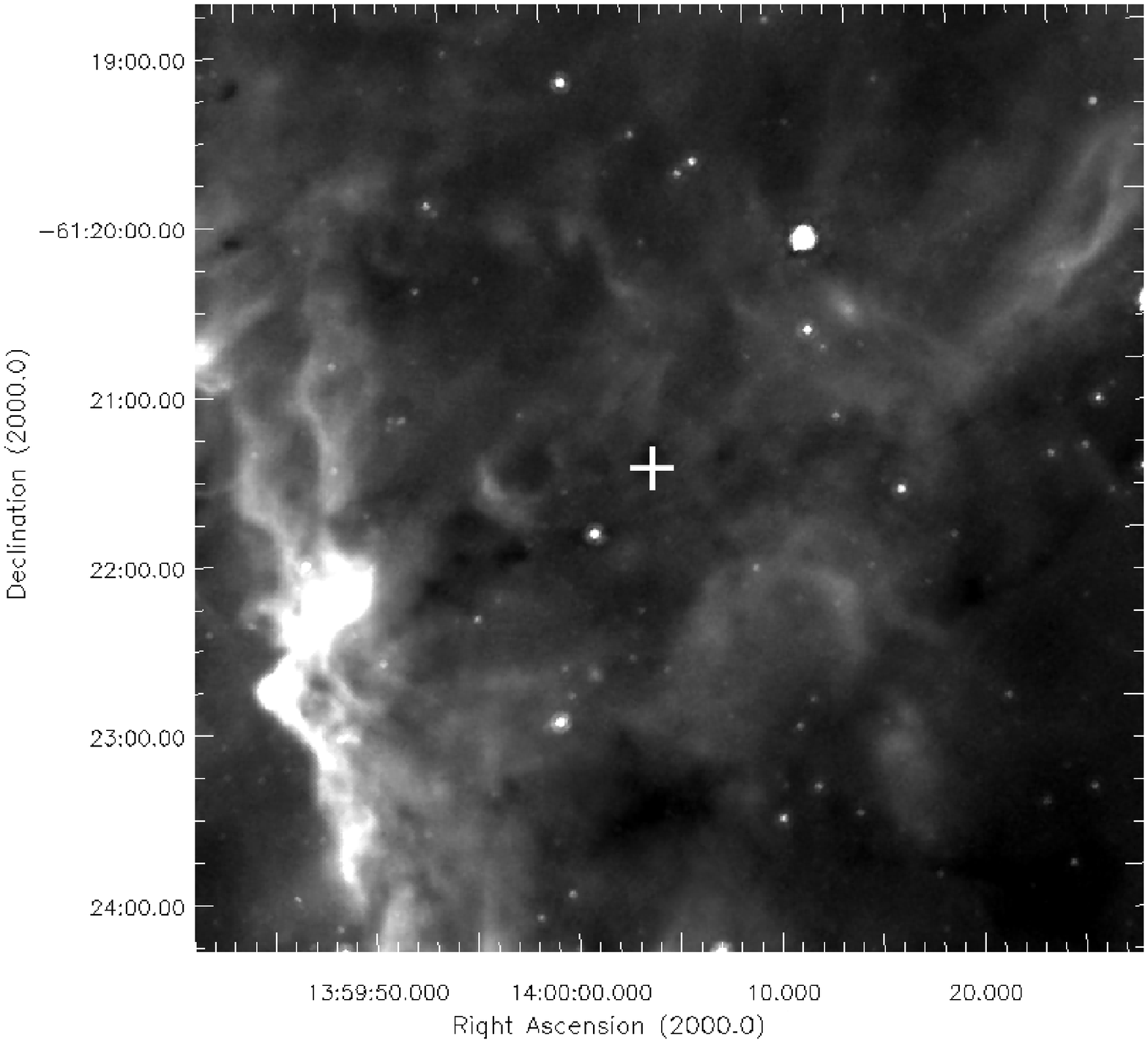}
\includegraphics[angle=0,width=85mm]{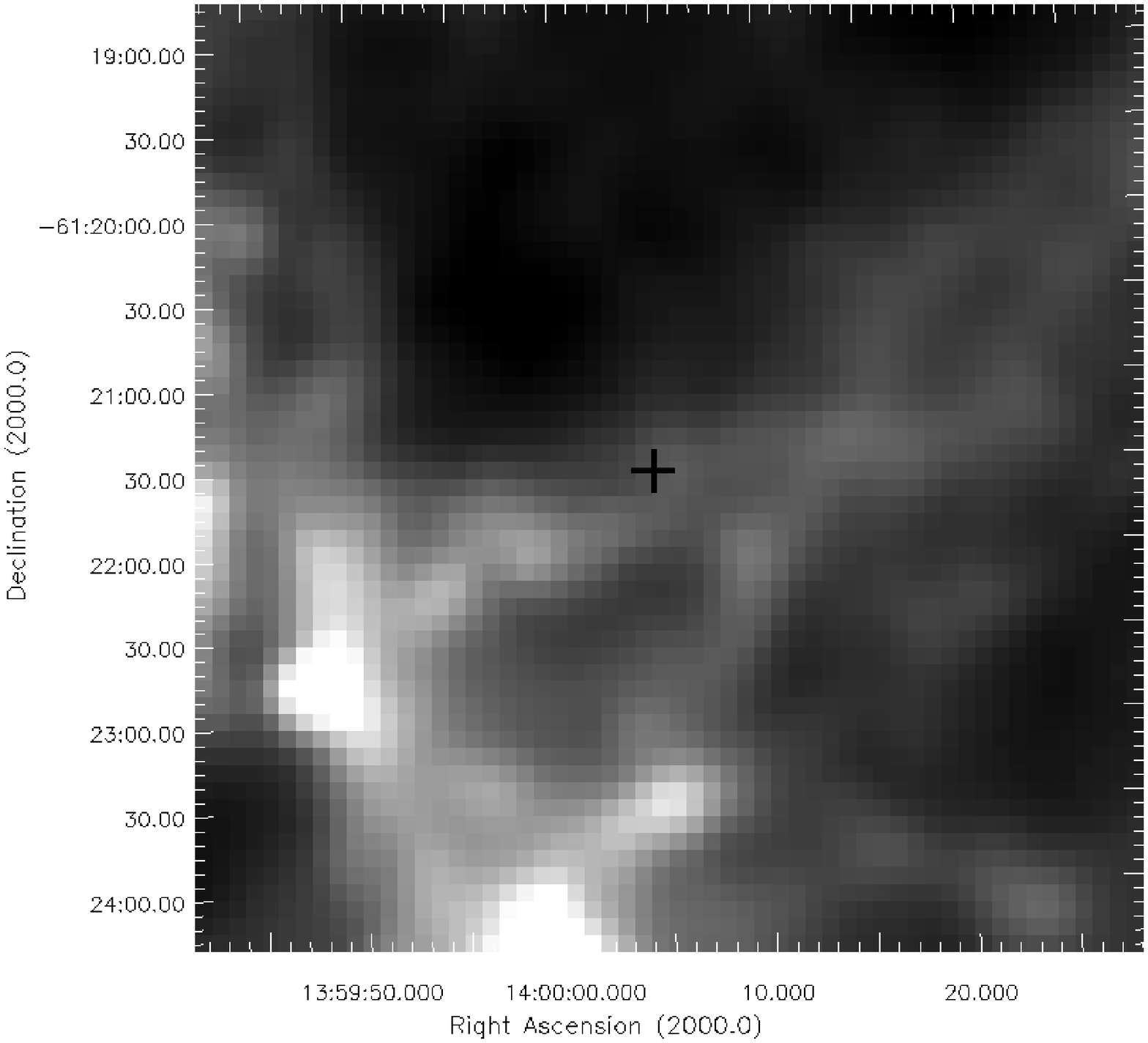}
\end{center}
\caption{G311.061+0.425. Left panel: 8~$\mu$m. Right panel: 250~$\mu$m.
The position of the IRDC is marked with a cross.
This is an example of a genuine \textit{Spitzer}-dark,
\textit{Herschel}-bright IRDC. However, it does not appear to contain
any dense cores. Only diffuse and filamentary
emission can be seen at 250~$\mu$m.} \label{nocore}
\end{figure*}

\section{RESULTS}\label{whatwedid}

\subsection{Classifying IRDCs}

The PF09 catalogue used \textit{Spitzer} 8-$\mu$m data to find 
11303 candidate IRDCs in the regions
$10\degr$$<$$l$$<$$65\degr$ and 
$295\degr$$<$$l$$<$$350\degr$ with
$|b|$$<$$1\degr$. They 
identified candidate IRDCs as connected structures with an
apparent mean 8-$\mu$m opacity 
greater than 0.35 and an apparent peak above 0.7. This
would correspond to a 
molecular hydrogen column density (N$_{H_{2}}$) detection 
threshold of a mean of $\sim 10^{22}$, 
and a peak of $\sim 2 \times 10^{22}$~cm$^{-2}$, 
respectively. Each
candidate IRDC had to be at least 4\arcsec\ in 
diameter. We refer to these regions as `\textit{Spitzer}-dark' regions.
To find the embedded cores (which PF09 termed fragments), they used 
apparent opacity contours with a step of 0.35. The number of local 
peaks between each consecutive level was then the number 
of fragments extracted. 

We present data here on the region from $l=300\degr$ to $l=330\degr$ with
$|b|$$<$$1\degr$. 
This region was chosen as it was the first large contiguous area to be 
covered in the Hi-GAL survey. It had also been observed in both the 
GLIMPSE and MIPSGAL surveys and was included in the PF09 survey. 
The region was therefore used to search for IRDCs and their cores. 
The PF09 catalogue contained 3171 \textit{Spitzer}-dark
candidate IRDCs in this region.
Identifying IRDCs in the MIR alone 
can cause problems. IRDCs appear as dark 
regions against the bright MIR background. However, there is no way
in the MIR of distinguishing between an area of low emission 
caused by absorption by an IRDC and an area of low emission 
caused by a local dip in the MIR background, sometimes
referred to as a `hole in the sky' \citep{stanke10}.
One way of identifying genuinely dense, cold regions is to look for 
emission in the far-infrared (FIR), where cold dust should emit
strongly. We label such regions as `\textit{Herschel}-bright' (referring
only to the longer wavelength \textit{Herschel} data).

\begin{figure*}
\begin{center}
\includegraphics[angle=0,width=85mm]{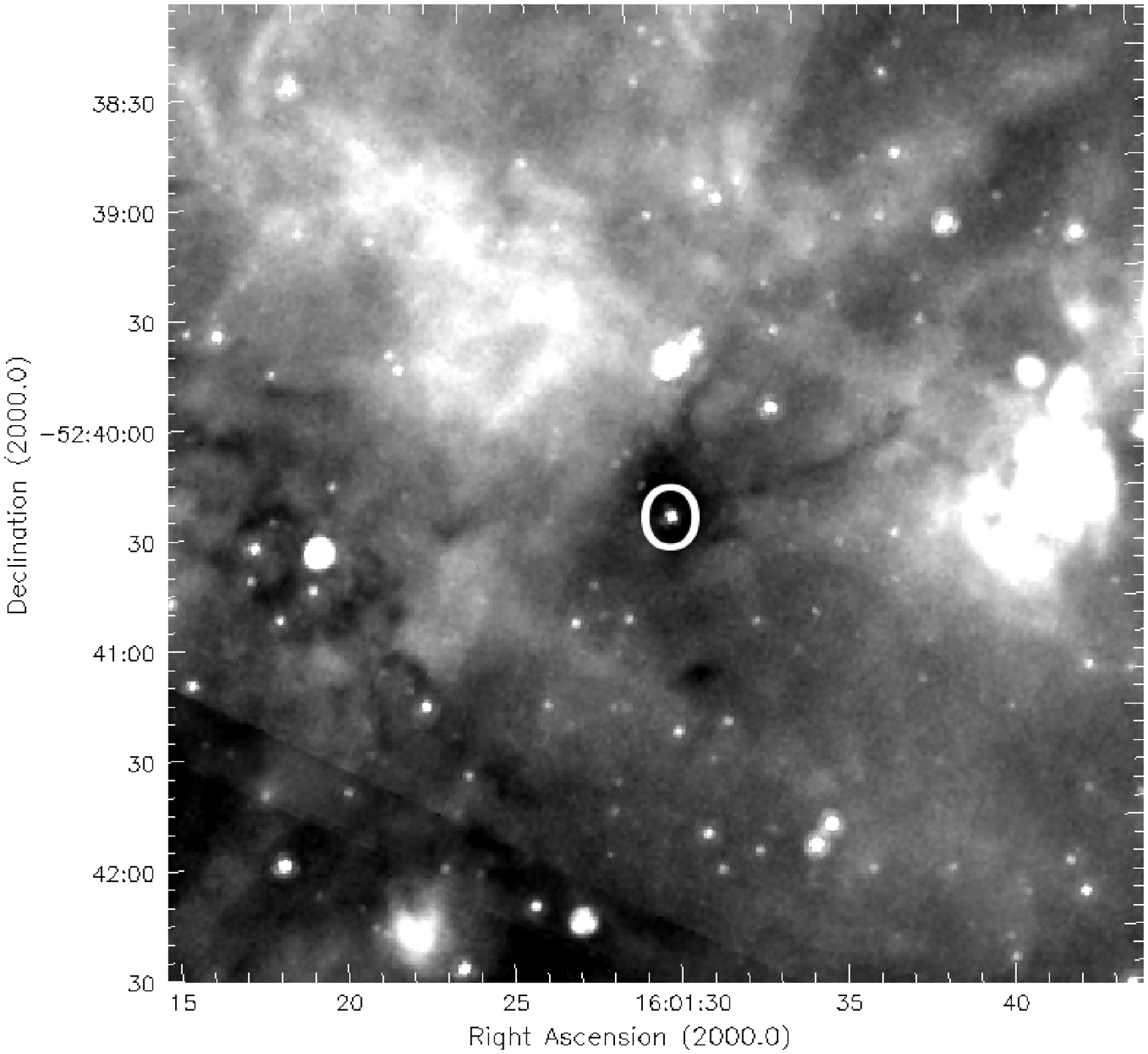}
\includegraphics[angle=0,width=85mm]{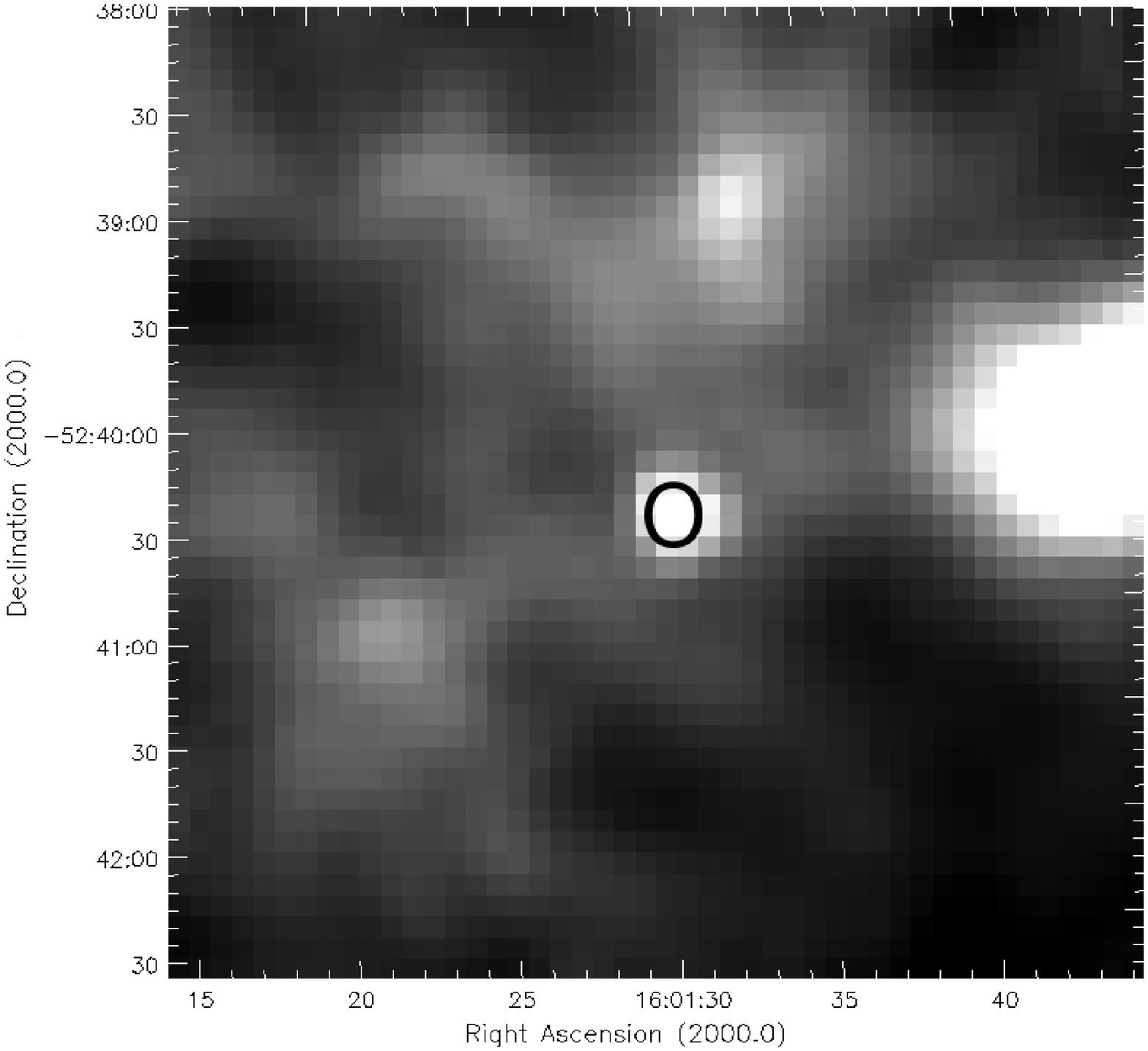}
\end{center}
\caption{G329.494+0.106. Left panel: 8~$\mu$m. Right panel: 250~$\mu$m.
This is an example of a genuine \textit{Spitzer}-dark and
\textit{Herschel}-bright IRDC, which contains an 8-$\mu$m point source. 
The point source is circled in both panels.}\label{eight}
\end{figure*}

We studied each of the 3171 objects within our search area
in the PF09 catalogue, using \textit{Spitzer} data at 8 and 24~$\mu$m,
and \textit{Herschel} data at 70, 160, 250, 350 
and 500~$\mu$m, to determine which of the objects were real IRDCs and
which were just local dips in the MIR emission. Some examples are shown in Figures 
\ref{realirdc}--\ref{eight}. Each candidate IRDC was viewed in a similar fashion to Figures 
\ref{realirdc}--\ref{eight} but at all six wavelengths.
Any object seen in absorption at 8~$\mu$m and seen simultaneously in
emission at 250, 350 and 500~$\mu$m was classed as 
both \textit{Spitzer}-dark and \textit{Herschel}-bright, and was classified as
a genuine IRDC. An example of a \textit{Spitzer}-dark and \textit{Herschel}-bright 
IRDC can be seen in Figure \ref{realirdc}. The full list of \textit{Herschel}-bright IRDCs is 
given in Appendix \ref{bright}.

At 70~$\mu$m a cloud may be expected to be seen in absorption, but  
less strongly than at 8~$\mu$m. Furthermore,  
the noise in the 70-$\mu$m data sometimes tended to
make it unclear whether the core was in
emission or absorption at 70~$\mu$m. Likewise,
a source's appearance at 160~$\mu$m depends heavily
on the temperature of the cloud. For these reasons, 
the 70- and 160-$\mu$m wavelengths were not used in the
initial classification process,
and were only used when the other wavelengths left
some ambiguity about a source's status.

Sensitivity limits on \textit{Herschel} mean that smaller, less
dense IRDCs are not likely to show enough emission to be classified as 
\textit{Herschel}-bright. We estimate that
any IRDC with a major axis less than 26\,\arcsec and a peak column density
less than 4$\times10^{22}$\,cm$^{-2}$ will not be visible in emission at
250, 350 or 500\,$\mu$m if it is situated in a region with a background
greater than 1300\,MJy/sr at 250\,$\mu$m. This accounts for approximately
20\,per\,cent of the PF09 objects. As it cannot be stated whether these objects are
emitting in the FIR or not, their true status is unknown and they remain
candidate IRDCs. Further details can be found in Appendix 
\ref{comp}.

In total, of the original 3171 objects in
the PF09 catalogue in our search area, we found only
1205 of them to be genuine IRDCs under our 
simultaneous \textit{Spitzer}-dark and \textit{Herschel}-bright definition. These objects 
are listed in Table~\ref{irdc} of Apeendix \ref{table}. We note  
that \textit{Herschel} is insensitive to $\sim$20\,per\,cent of the candidates, see Appendix 
\ref{comp}.

This method relies upon human interpretation of the data. To estimate the error-bars introduced 
by this method a second person observed all the candidates in a circular 
region with a radius of 0.5\,\degr. They classified each candidate as \textit{Herschel}-bright 
or \textit{Herschel}-dark as before. Of the 107 candidate IRDCs in this region, only three were 
classified differently by the second person. 

This implies that all previous catalogues of IRDCs, 
based solely on MIR data, may have over-estimated the total
number of IRDCs in the Galaxy. In this region the number has been
over-estimated by up to a factor of $\sim$1.7--2.6. Similar factors
might be expected elsewhere.
\citet{jackson08} carried out a `reliability' test on the 
IRDC catalogue of \citet{simon06c}, and found values
ranging from $\sim$50\,per\,cent to $\sim$100\,per\,cent for the fraction of genuine
IRDCs in the catalogue, depending on the contrast level in the MIR.
In other words, they found up to a factor of $\sim$2 over-estimate in
the number of genuine IRDCs -- consistent with our
results.

The discovery that only 1205 of the candidate IRDCs, in
a catalogue based on MIR data, turn out 
to be \textit{Herschel}-bright (see upper part of 
Table~\ref{finalnumbers}) has ramifications
for all such catalogues based on MIR data alone.
The total number of candidate IRDCs in the catalogues
of \citet{simon06c} and PF09 is $\sim$11000 in each.
If our observed ratio is consistent throughout these
catalogues, then the total number of genuine IRDCs in each
may be as low as $\sim$4000--6000 (c.f. \citealt{jackson08}).
This has consequences for calculations of the
total number of IRDCs in the Galaxy.

\subsection{Cores within IRDCs}

Each cloud was examined at 250~$\mu$m. Some IRDCs were
seen to have relatively simple structures,
while others had more complex structures.
The 250-$\mu$m band was chosen as 
this had the best resolution of the three wavelengths where 
the IRDCs were predicted to be seen in emission.
Each cloud was examined for evidence that one or more dense 
cores had formed within the cloud.

A gaussian profile was fitted towards the peak of the intensity map of each core at 
250~$\mu$m. 
Our criteria for classification as \textit{Herschel}-bright specified that a candidate IRDC 
had to be seen in emission at 250, 350 and 500\,$\mu$m. The emission from each of our 
\textit{Herschel}-bright IRDCs was therefore much greater than its surrounding background. As such, 
no background subtraction was needed when fitting the gaussian profiles.

The gaussian fitted at 250~$\mu$m was used 
to determine the full-width at half-maximum (FWHM) of each core. 
Some IRDCs could be fitted with more than one gaussian, indicating
the presence of more than one dense core. A total of
972 IRDCs fell into the category of having one or more dense
cores within them.

In the case of 215 IRDCs, cloud emission was seen in the 
FIR at 250~$\mu$m, but there was no discernible 250-$\mu$m peak.
These IRDCs were deemed not to have any dense cores within
them. For 18 IRDCs the data were found 
to be excessively `stripy' at 250~$\mu$m, 
resulting in no gaussian being able to be fitted.
These were discarded.
These 233 objects, to which no gaussian could be fitted, 
were omitted from further consideration.
This left us with a `clean' sample of 972 IRDCs that contain
one or more dense cores, on which we concentrate for the
remainder of the paper.

\subsection{Protostars within IRDC cores}

The 972 IRDCs containing one or more dense
cores were studied closely in the 8-$\mu$m data
for evidence of an 8-$\mu$m point source.
The presence or lack of an 8-$\mu$m point source within 
a core is most likely to be indicative of the presence or 
absence of an embedded protostar. Hence, this is an
indication of the evolutionary status of the core. 
More evolved cores, namely those already undergoing star formation,
are more likely to have an 8-$\mu$m point source within them. 

Every core in all of the 972 IRDCs was searched for an embedded 8-$\mu$m
point source. 
An 8-$\mu$m point source was defined as a compact, roughly circular source with
an 8-$\mu$m peak of greater than 3$\sigma$, where $\sigma$ is the noise level of the data and was 
defined as the standard deviation in flux towards the edge of
each region. The peak of the point source had to be
within a radius from the centre of the core (defined at 250~$\mu$m) equal to the FWHM of 
the core at 250~$\mu$m. It should be noted that, as no distance
information is available for the majority of these IRDCs, it is possible
that the 8\,$\mu$m point sources noted here are, in some cases, not
associated with the IRDC itself but are instead foreground stars
contaminating the field of view (see also \citealt{lumsden02}). 
653 out of 972 IRDCs (67\,per\,cent)
were found to have at least one 8-$\mu$m point source embedded in 
one or more of their constituent cores.
This left 319 IRDCs with no 8-$\mu$m point source. 

These 319 IRDCs without 8-$\mu$m point sources were 
searched for a 24-$\mu$m point source.
The search was carried out
using the same criteria as when looking for an 8-$\mu$m
point source.  
Of the 319 IRDCs, 149 were found to have one or more
24-$\mu$m point sources. 

In summary, we found
a total of 972 IRDCs that contained one or more discernable cores
at 250~$\mu$m. Of these, 653 have an 8-$\mu$m point source.
Of the IRDCs with no 8-$\mu$m point source,
a further 149 have a 24-$\mu$m point source.
We designate the remaining 170 as starless IRDCs.
These objects could
be the high-mass equivalents of low-mass starless cores. The total 
number of IRDCs in each category is summarised in 
Table~\ref{finalnumbers}.

\section{Discussion} \label{lifetimes}

\subsection{Statistics}

\begin{table*}
\begin{center}
\caption{IRDC statistics. The upper part of the Table lists the number and 
percentage of candidates in the PF09 catalogue that were found 
to be \textit{Herschel}-bright. The lower part of the Table
refers to the sample of 972 IRDCs that contained one or more dense 
cores at 250~$\mu$m, and lists those with embedded 8-$\mu$m sources,
those without 8-$\mu$m sources that contain 24-$\mu$m sources,
and those with neither 8-$\mu$m nor 24-$\mu$m sources. }
\label{finalnumbers}
\begin{tabular}{lccc} \hline
Source & Equivalent Class of & Number of & Percentage\\ 
Type  & \protect\citet{chambers09} & IRDCs &\\
 & & &\\ \hline 
 & & &\\
PF09 sample & & 3171 & 100 \\
Herschel-bright & & 1205 & 38 \\
 & & & \\ \hline \hline
 & & & \\
IRDCs with cores & & 972 & 100 \\
IRDCs with 8-$\mu$m source & Red Core & 653 & 67 \\
IRDCs with 24-$\mu$m source only & Active Core & 149 & 15 \\
No 8- or 24-$\mu$m source & Quiescent Core & 170 & 18\\
 & & & \\ \hline
\end{tabular}
\end{center}
\end{table*}

The relative numbers of IRDCs with no embedded MIR
point sources, compared to those with 8- and 24-$\mu$m
point sources (see lower part of Table~\ref{finalnumbers}), 
can be used for comparison with previous findings.
\citet{chambers09} labelled IRDCs with no
embedded MIR point sources as quiescent, those with a 24-$\mu$m 
point source as active, and those with an 8-$\mu$m point source
as red cores.
They had a sample of 190 candidate IRDCs, and found
$\sim$21\,per\,cent to be red,
$\sim$25\,per\,cent to be active and 
$\sim$54\,per\,cent to be quiescent.
These percentages can be compared to the last three lines in
Table~\ref{finalnumbers}.

It can be seen that there is a much larger fraction of quiescent 
IRDCs in the Chambers sample than in ours. This could be due
to the effect of some fraction of their initial sample of
candidate IRDCs not being genuine, as they had no FIR data. Interestingly, though, because they 
see 46\,per\,cent of their candidate IRDCs having other star formation tracers,
then only a maximum of 54\,per\,cent ($\pm$5\,per\,cent) of their candidate IRDCs could be 
false, compared to up to 62\,per\,cent ($\pm$1\,per\,cent) in our full sample
(although note that some of their associations could be chance alignments).
However, \citet{chambers09} have a much smaller sample of
cores. Additionally, \citet{chambers09} use the 
\citet{simon06c} catalogue of candidate IRDCs. The \citet{simon06c} catalogue was based on 
\textit{MSX} data and used different criteria to PF09 when finding candidate IRDCs. This makes a
detailed comparison between this work and that of \citet{chambers09} difficult. The errors
quoted in brackets here are simply the poisson ($\sqrt n$)
errors, and there may also be systematic effects at work.
Hence, these two numbers are roughly consistent.

\citet{parsons09} cross-matched the candidate IRDCs of 
\citet{simon06c} with SCUBA 850-$\mu$m emission, and found that
25\,per\,cent of the candidate IRDCs were not seen in emission at 850~$\mu$m.
Of those that were detected by SCUBA, which also appeared in
the GLIMPSE survey, they found that 70\,per\,cent had embedded 24-$\mu$m
point sources, and 30\,per\,cent did not. They made no distinction
between those 24-$\mu$m sources that also had 8-$\mu$m point sources
and those that did not. Hence their 30\,per\,cent needs to be compared
to the last line in Table~\ref{finalnumbers} (18\,per\,cent), and their
70\,per\,cent should be compared to the sum of the two lines above that 
in Table~\ref{finalnumbers} (82\,per\,cent). Given the very different
sizes of the samples (the simple poisson $\sqrt n$ errors on
their numbers are $\sim\pm$10\,per\,cent), these are also consistent
with the percentages we find.

\subsection{Relative lifetimes}

A simple evolutionary picture for a core within an IRDC is given by \citet{chambers09}. This 
entails a quiescent, starless core evolving first into a core 
with an embedded 24-$\mu$m point source. This phase could be 
equated theoretically with the main accretion phase for massive 
protostar formation. Subsequently the core would evolve into a 
core with an embedded 8-$\mu$m point source, indicating
the beginnings of an HII region starting to form.
If each starless IRDC core evolves into a corresponding 
star-forming core with one or more embedded 24-$\mu$m point sources,
and each of these evolves into a core with the same number of embedded 
8-$\mu$m point sources,
then we can equate the above statistics with theoretical lifetimes,
and produce statistical lifetimes for each of the stages in 
Table~\ref{finalnumbers}.

One clear result that then comes from Table~\ref{finalnumbers} is 
that only about one-fifth of IRDC cores do not contain embedded point sources.
If this corresponds to the `starless' phase of IRDCs, then
we would conclude that a typical IRDC core only spends around
one-fifth of its lifetime without any seed of a protostar 
within it. \citet{chambers09} assume that the 
accretion timescale of high mass star formation is
equivalent to the amount of time that an IRDC core will exist in  
the active phase (i.e. has a 24-$\mu$m point source, but no
8-$\mu$m point source). 
We can use a canonical value for the accretion timescale 
of $\sim$2$\times$10$^5$~years \citep{zinnecker07} 
as the lifetime for the active phase. 

Hence we would calculate, from the similarity of the percentages
in the last two lines of Table~\ref{finalnumbers}, that the
lifetime of the quiescent, or starless, phase of IRDCs is also
$\sim$2$\times$10$^5$~years.
Similarly, the lifetime of the phase with an embedded 8-$\mu$m source
(the red stage in Chambers' nomenclature)
would then be $\sim$6$\times$10$^5$~years and the entire IRDC
lifetime (after core formation) would be $\sim$10$^6$~years.
\citet{chambers09} find $\sim$4$\times$10$^5$~years for 
the quiescent phase, also by assuming 2$\times$10$^5$~years 
for the active phase. 

The 8-$\mu$m emission is believed to arise from very hot, very
small grains, known as polycyclic aromatic hydrocarbons (PAHs;
\citealt{chambers09}).
A typical PAH is so small that a single high-energy photon can 
interact with a PAH and raise its temperature to a few hundred 
degrees for a short period of time, before it cools again by 
re-emitting a photon in the MIR. Such high-energy photons are
presumed to ionise the surrounding material and create 
hyper-compact and ultra-compact HII
(HCHII and UCHII) regions. Hence the phase where an IRDC
contains an embedded 8-$\mu$m point source should roughly 
correlate with the combined HCHII and UCHII phases. The 
UCHII lifetime is not well known, although recent estimates put
it at several~$\times$~10$^5$~years (e.g. \citealt{kaper11}),
consistent with our estimate.
Similarly, \citet{mckee02} look at stars in typical regions of
high mass star formation and find timescales of
a few~$\times$10$^5$~years for the combined HCHII and UCHII phases. 
This is also consistent with our estimated lifetime of the phase
in which an IRDC has an embedded 8-$\mu$m point source.

\section{Conclusions}

3171 candidate IRDCs were catalogued from their MIR absorption in 
\textit{Spitzer} data (\textit{Spitzer}-dark regions). We found 1205 
which are \textit{Herschel}-bright. The other objects 
may be simply minima in the IR background. This suggests that IRDC searches based solely on MIR 
data may over-estimate the total number of IRDCs in the Galaxy by up to a factor of $\sim$2.

972 of the 1205 \textit{Herschel}-bright IRDCs have one or more discernible peaks at
250~$\mu$m, indicating the formation of dense cores within these IRDCs.
The \textit{Spitzer} data were then examined to see whether the
IRDCs with cores contained either an 8- or a 24-$\mu$m point source. 
653 are seen to harbour one or more 8-$\mu$m point sources, and of
the remainder, a further 149 contained one or more 24-$\mu$m point 
sources.

We equated the presence of a 24-$\mu$m point source to the
typical accretion timescale for high-mass stars
of $\sim$2$\times$10$^5$~years and hence derived a timescale
for the starless IRDC core phase also of 
$\sim$2$\times$10$^5$~years.
We equated the presence of an 8-$\mu$m point source to the
combined HCHII and UCHII phase, and derived a timescale of $\sim$6$\times$10$^5$~years for this 
stage. A total lifetime for IRDCs with dense cores of $\sim$10$^6$~years was thus derived.

\section*{Acknowledgements}

LAW acknowledges STFC studentship funding.
SPIRE was developed by a consortium of institutes led by 
Cardiff University (UK) and including Univ. 
Lethbridge (Canada); NAOC (China); CEA, LAM (France); IFSI, Univ. 
Padua (Italy); IAC (Spain); 
Stockholm Observatory (Sweden); Imperial College London, RAL, 
UCL-MSSL, UKATC, Univ. Sussex 
(UK); and Caltech, JPL, NHSC, Univ. Colorado (USA). This development 
was supported by national 
funding agencies: CSA (Canada); NAOC (China); CEA, CNES, CNRS (France); 
ASI (Italy); MCINN 
(Spain); SNSB (Sweden); STFC (UK); and NASA (USA). 
PACS was developed by a consortium of institutes led by MPE (Germany) 
and including UVIE
(Austria); KU Leuven, CSL, IMEC (Belgium); CEA, LAM (France); 
MPIA (Germany); INAF-IFSI/OAA/OAP/OAT, LENS, SISSA (Italy); IAC
(Spain). This development was supported by the
funding agencies BMVIT (Austria), ESA-PRODEX (Belgium), 
CEA/CNES (France), DLR (Germany),
ASI/INAF (Italy), and CICYT/MCYT (Spain). 
HIPE is a joint development by the \textit{Herschel} Science Ground 
Segment Consortium, consisting of ESA, the NASA \textit{Herschel} 
Science Center, and the HIFI, PACS and 
SPIRE consortia.
This work is also based, in part, on observations made with the 
\textit{Spitzer} Space Telescope, which
is operated by the Jet Propulsion Laboratory, California Institute 
of Technology under a contract with NASA.

\bibliographystyle{mnras.bst}
\bibliography{bib}

\appendix
\include{cat}\label{table}
\include{cat_comp}

\end{document}

%% file: cat.tex
\onecolumn

\section{List of Herschel-bright IRDCs}\label{bright}

\begin{center}
\begin{longtable}{lccccc} 
\caption{The 1,205 IRDCs which were found to be \textit{Herschel}-bright. Column 1 states the IRDC name as it
appears in PF09. Columns 2 and 3 are the position of the IRDC. 
Columns 4 and 5 give details on whether the object contains a
MIR source. Column 6 gives additional information about the 
cloud, including whether it contains a visible core and whether it
is isolated at 160 and 500\,$\mu$m. } \label{irdc}\\

\hline IRDC Name & Right Ascension & Declination & 8\,$\mu$m & 24\,$\mu$m & Additional comments \\ 
 & 2000.0 ($^{\circ}$) & 2000.0 ($^{\circ}$) & source? & source? & \\ \hline \endfirsthead

\hline IRDC Name & Right Ascension & Declination & 8\,$\mu$m & 24\,$\mu$m & Additional comments \\ 
& 2000.0 ($^{\circ}$) & 2000.0 ($^{\circ}$) & source? & source? & \\ \hline \endhead

\hline \textit{continued on next page}\\ \endfoot
\hline \\ \endlastfoot

300.345$-$0.357  & 187.13 &  $-$63.11  &  Yes  &    &  \\ 
300.356$-$0.358  & 187.16 &  $-$63.12  &  Yes  &    &  \\ 
300.596+0.098  & 187.77 &  $-$62.68  &  Yes  &    &  \\ 
300.743+0.036  & 188.08 &  $-$62.75  &  Yes  &    &  \\ 
300.746+0.550  & 188.16 &  $-$62.24  &  Yes  &    &  \\ 
301.131$-$0.274  & 188.88 &  $-$63.09  &  Yes  &    &  \\ 
301.206$-$0.193  & 189.06 &  $-$63.01  &  No  &  No  &  Confused at 500\,$\mu$m\\ 
301.283$-$0.242  & 189.22 &  $-$63.07  &  Yes  &    &  \\
\end{longtable} 
\end{center}

\twocolumn

%% file: cat_comp.tex
\section{Testing the Catalogue Completeness}\label{comp}

In the main body of this paper, we search for IRDCs within the area $l=300-330\,\degr$,
$|b|<1\,\degr$ using the catalogue of PF09. They used \textit{Spitzer} 
8-$\mu$m data to identify 11303 candidate IRDCs in the Galactic Plane. 3171 of these lie 
in our search area. To ascertain how many IRDCs may have been missed due to sensitivity limits we modelled several IRDCs
and placed them in the Hi-GAL data in regions with differing backgrounds. The modelled IRDCs were
observed in the same manner as the original candidates. The dimensions and flux levels of the smallest
visible IRDC were determined.

To model the IRDCs we used \textsc{Phaethon} \citep{stamatellos03, stamatellos05,
stamatellos10}. \textsc{Phaethon} is a 3D Monte Carlo radiative transfer code and has been used
previously to model IRDCs \citep{stamatellos10, wilcock11}. The code uses 
luminosity packets to represent the ambient radiation field in the system. These packets are 
injected into the system where they interact (are absorbed, re-emitted or scattered) with it 
stochastically. The ambient radiation field is taken to be a multiple of a modified version of 
the \citet{black94} interstellar radiation field (ISRF), which gives a good approximation to the
radiation field in the solar neighbourhood. The input variables of the code are the density 
profile, the strength of the ambient radiation field, the dust properties of the system, the 
size of the core and its geometry.

IRDCs were created with three different radii (0.2, 0.4 and 0.7\,pc) and two different peak 
column densities (the detection threshold of PF09, $2\times10^{22}$\,cm$^{-2}$, and 
$4\times10^{22}$\,cm$^{-2}$). Using the mean parameters of IRDCs modelled by \citet{20mo},
we place our model IRDCS at a distance of 3.1\,kpc and use a 
surrounding ISRF of 3.2 times the \citet{black94} radiation field. As most IRDCs do not appear
spherical, they were modelled with a flattened geometry which has a density profile given by:
\begin{equation}
n \left( r, \theta \right) = n_0 \left( {\rm H}_2 \right) \frac{1 + A \left( \frac{r}{R_0} \right) ^2 
\left[ \sin (\theta) \right] ^2 }{\left[ 1 + \left( \frac{r}{R_0} \right) ^2 \right] ^2} ,
\end{equation}
where \textit{r} is the radial distance, $\theta$ is the polar angle and $R_0$ is the flattening radius 
(i.e. the radial distance for which the central density is approximately constant). \textit{n}$_0$(H$_2$) 
is the central density, which is controlled as an input variable. \textit{A} is a factor that controls the 
equatorial to polar optical depth ratio and determines how flattened the core is. This was set at 2.5 and corresponds to an
aspect ratio of $\sim1:7$. 
$R_0$ was one tenth of the maximum radius. The dust opacity at 500\,$\mu$m used was
0.03\,cm$^2$\,g$^{-1}$ \citep{ossenkopf94}.

\begin{figure*}
\begin{center}
\includegraphics[width=55mm,angle=-90]{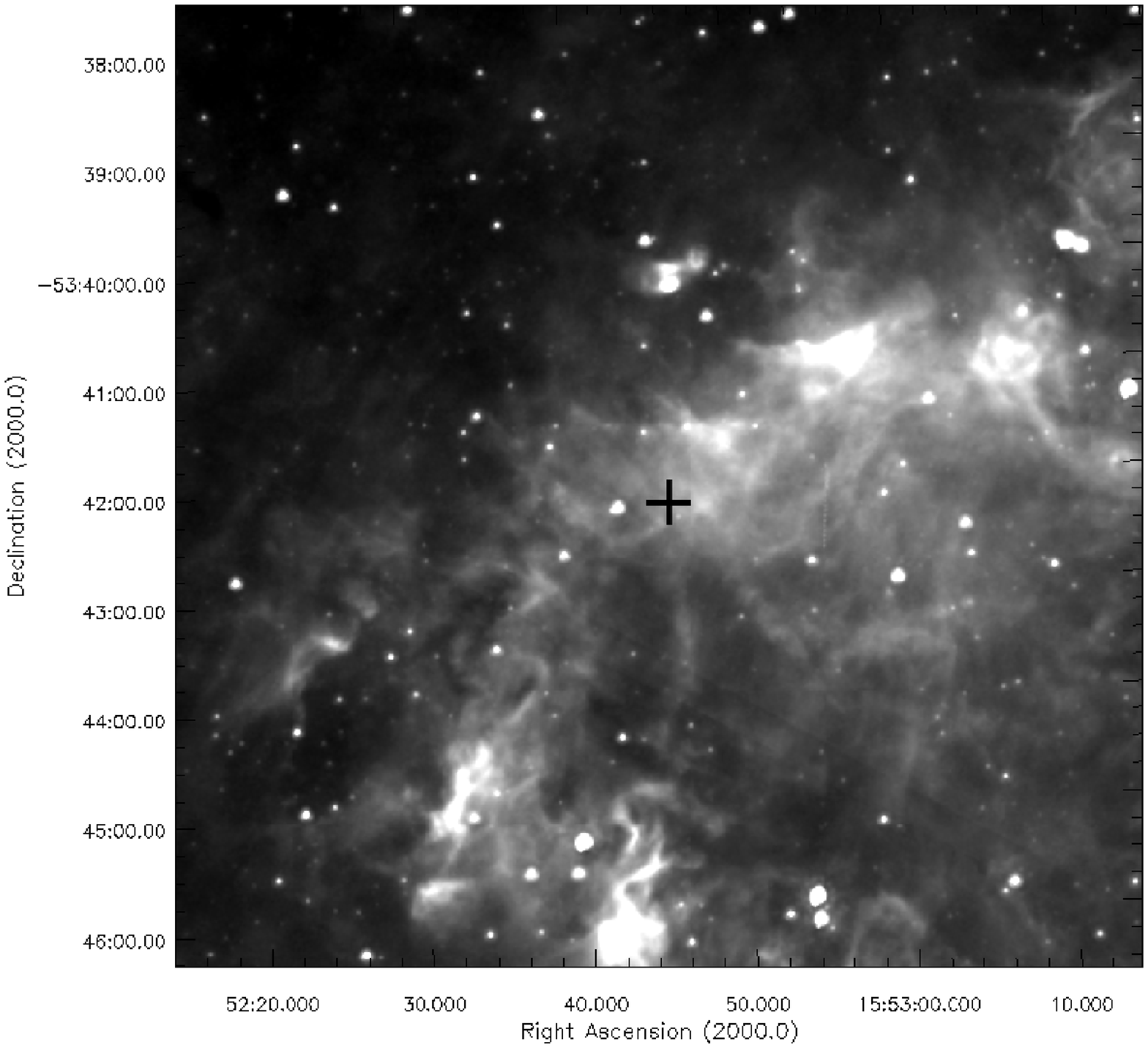}
\includegraphics[width=55mm,angle=-90]{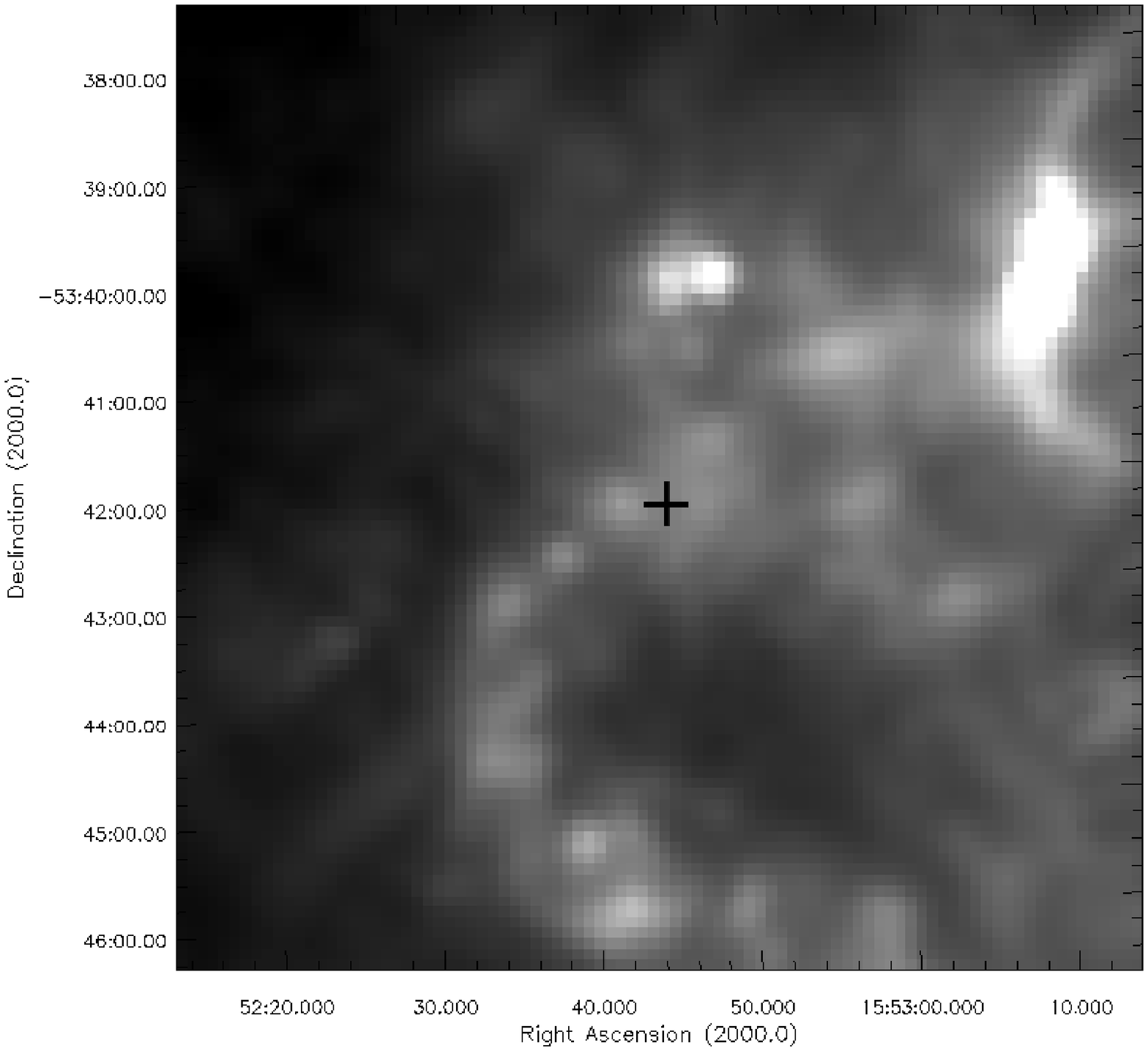}
\end{center}
\caption{Position A, shown without the addition of any modelled IRDCs, at 8\,$\mu$m (left) and
250\,$\mu$m (right). A cross marks the point where the IRDCs are added. } \label{da_a}
\end{figure*}

\begin{figure*}
\begin{center}
\includegraphics[width=55mm,angle=-90]{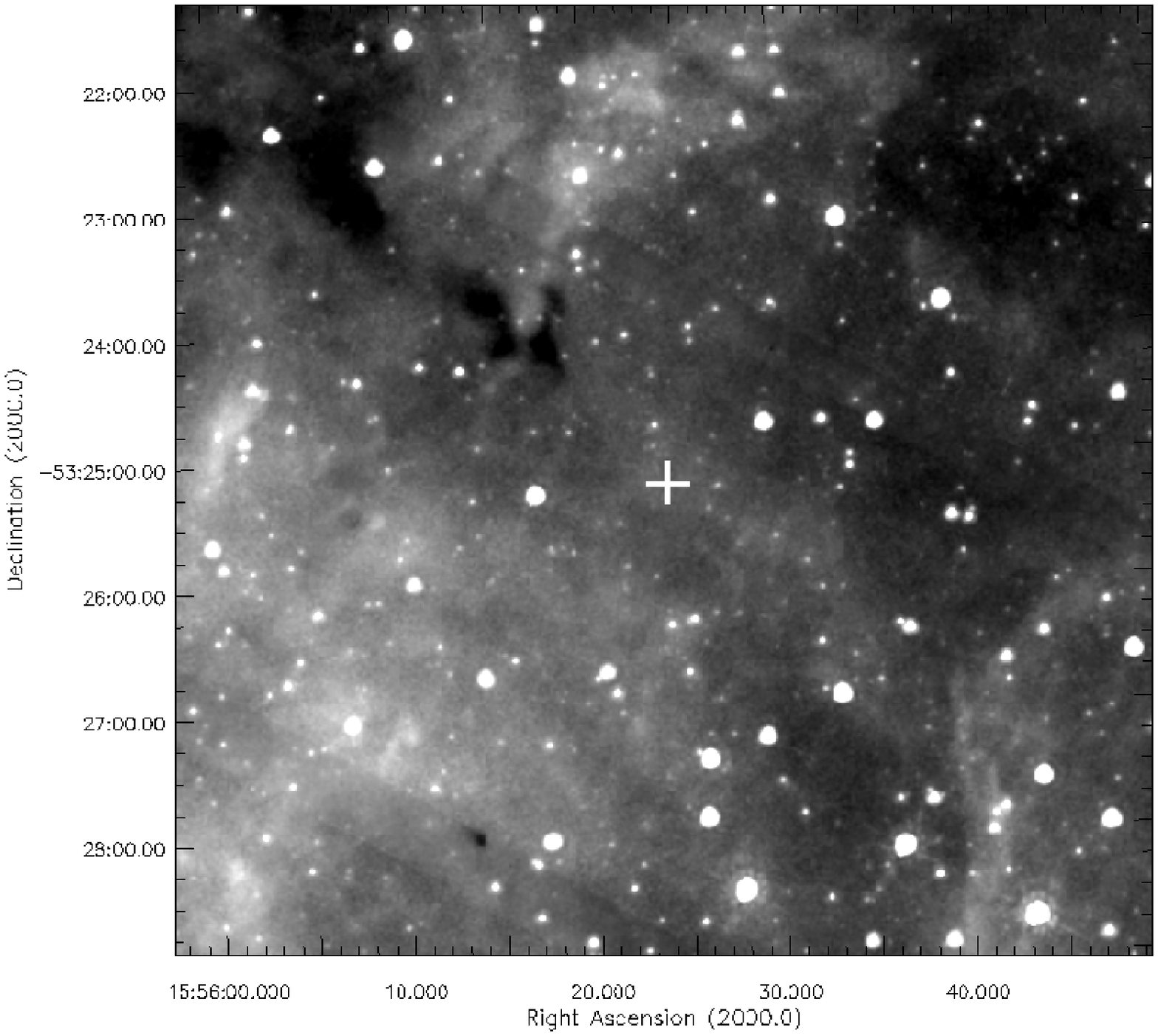}
\includegraphics[width=55mm,angle=-90]{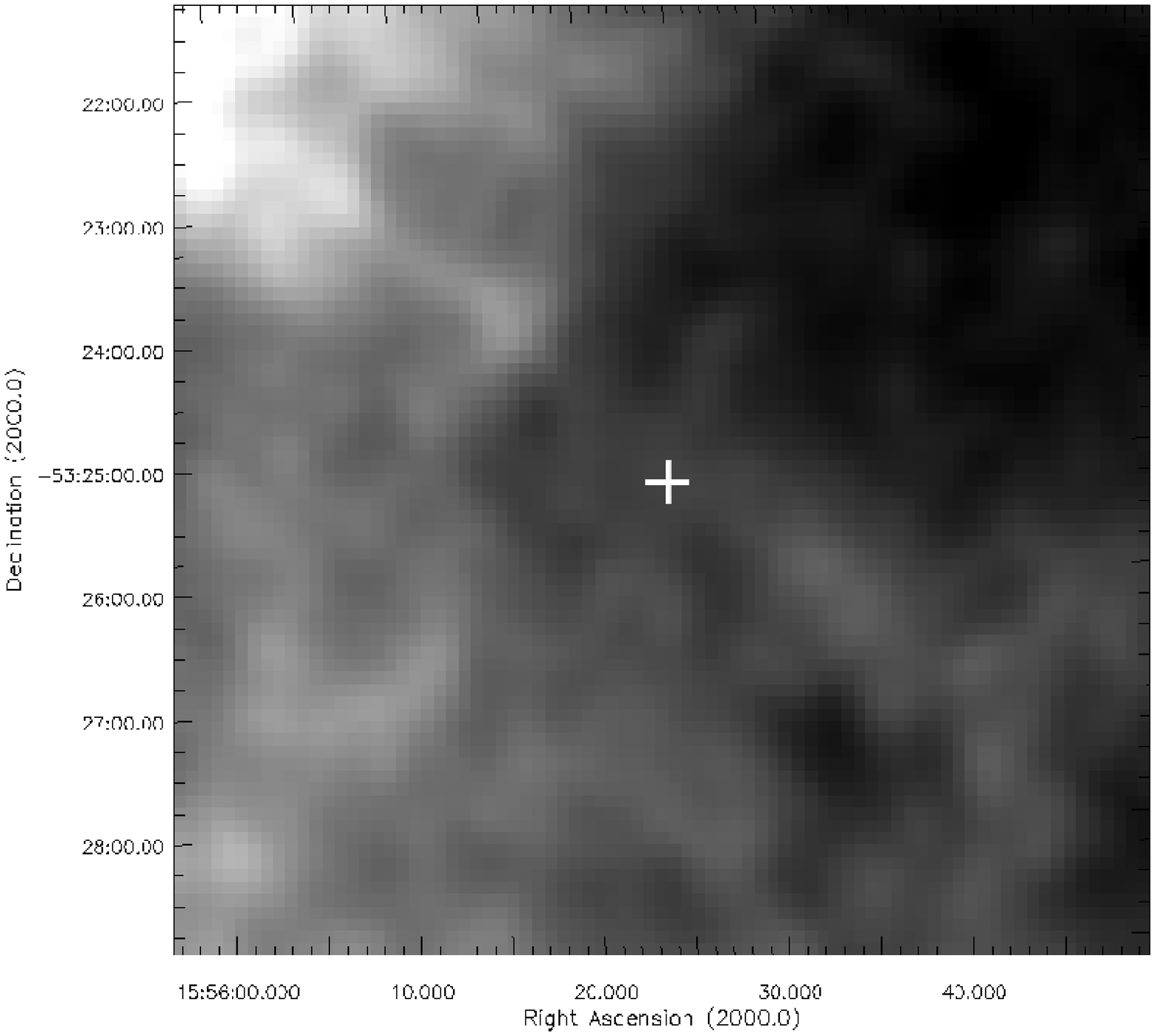}
\end{center}
\caption{Position B, shown without the addition of any modelled IRDCs, at 8\,$\mu$m (left) and
250\,$\mu$m (right). A cross marks the point where the IRDCs are added.} \label{da_b}
\end{figure*}

\begin{figure*}
\begin{center}
\includegraphics[width=55mm,angle=-90]{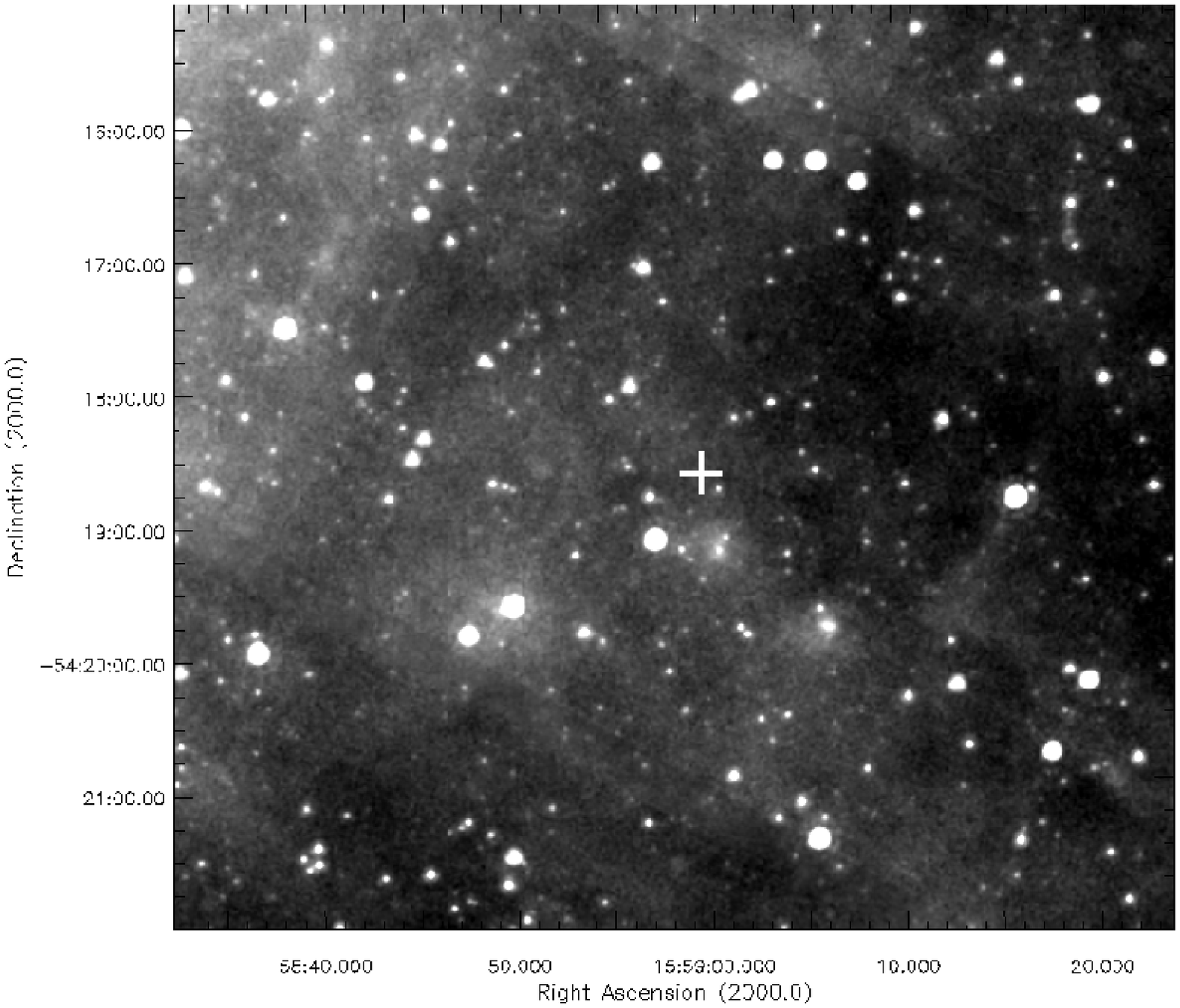}
\includegraphics[width=55mm,angle=-90]{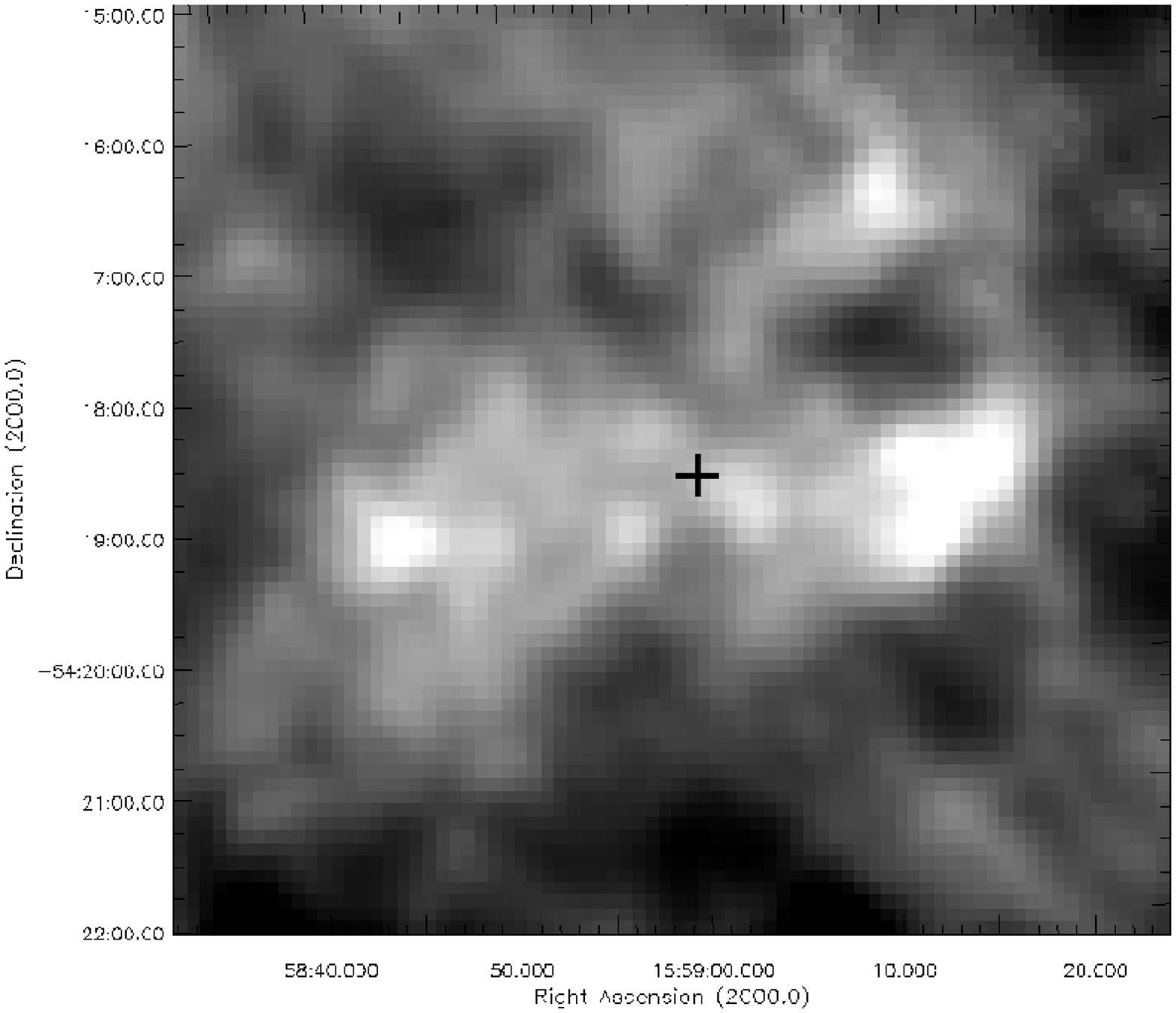}
\end{center}
\caption{Position C, shown without the addition of any modelled IRDCs, at 8\,$\mu$m (left) and
250\,$\mu$m (right). A cross marks the point where the IRDCs are added.} \label{da_c}
\end{figure*}

\begin{figure*}
\begin{center}
\includegraphics[width=55mm,angle=-90]{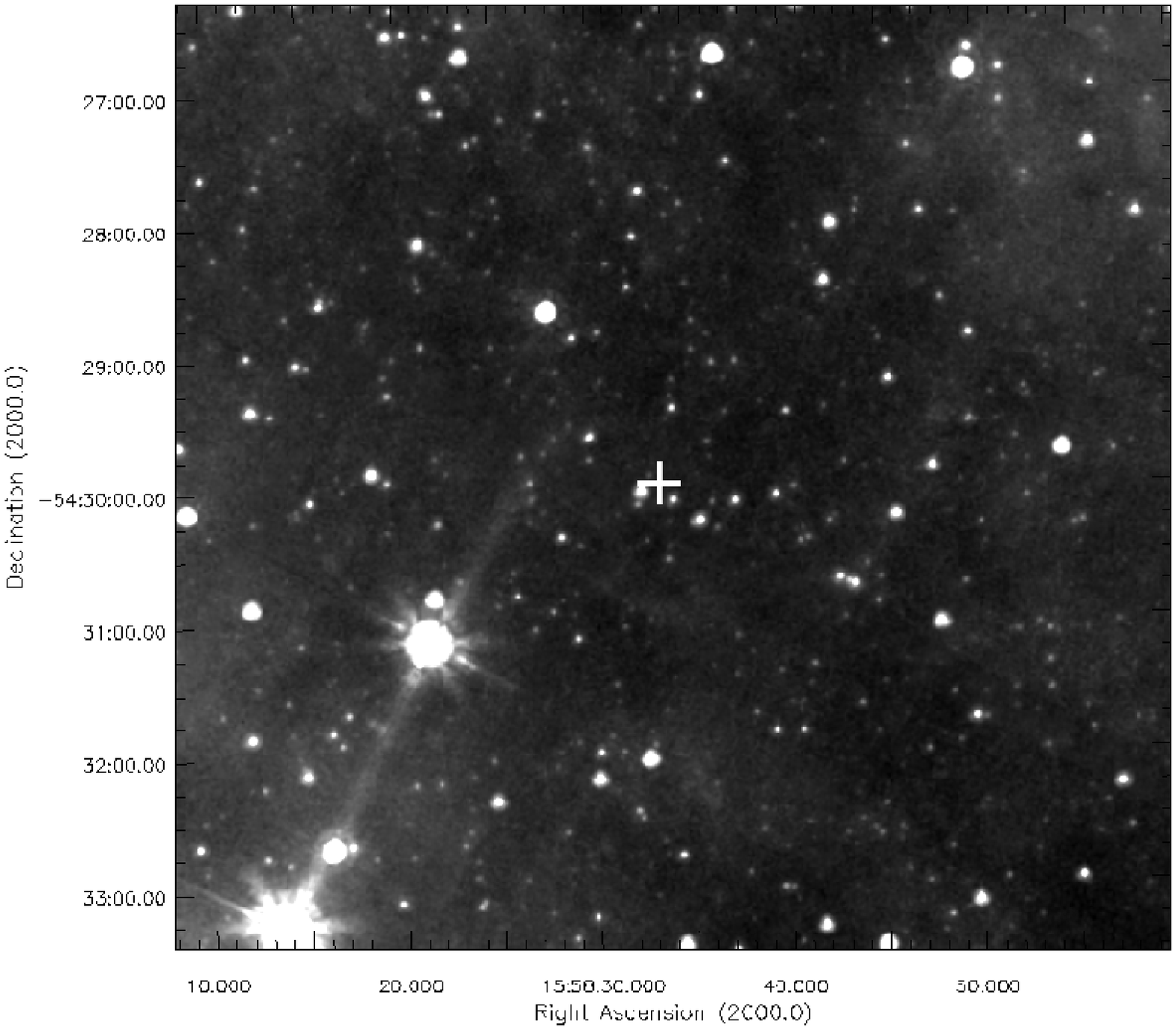}
\includegraphics[width=55mm,angle=-90]{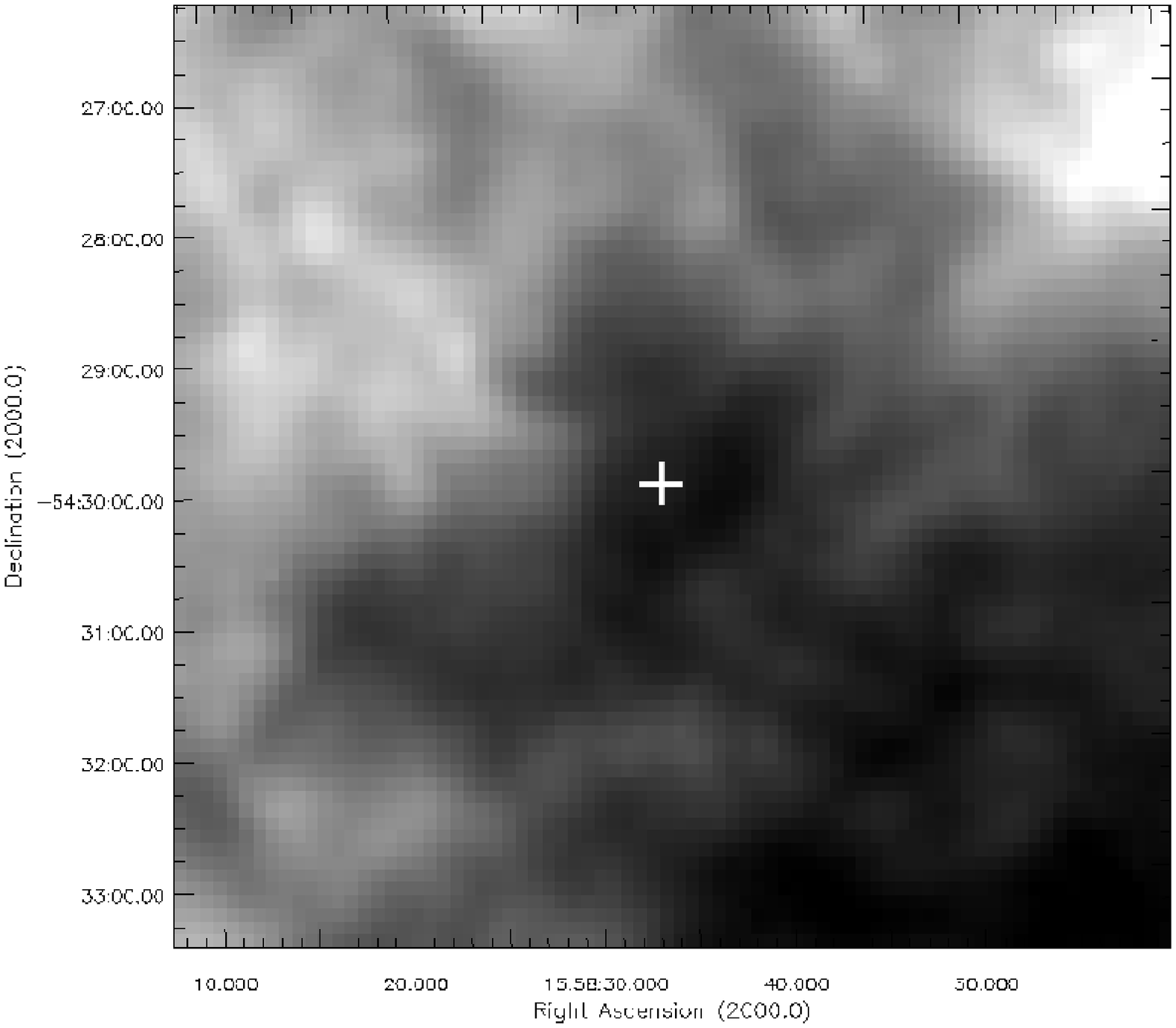}
\end{center}
\caption{Position D, shown without the addition of any modelled IRDCs, at 8\,$\mu$m (left) and
250\,$\mu$m (right). A cross marks the point where the IRDCs are added.} \label{da_d}
\end{figure*}

Each of the six IRDCs was convolved with the telscope beam at each wavelength and placed into the 
Hi-GAL data in four positions. The locations selected 
were typical areas within the Hi-GAL field that did not contain any candidate IRDCs and were
chosen to cover a range of different background levels. These were: Position A 
($l=327.829$\,\degr, $b=+0.17$\,\degr), a confused region near the centre of the Galactic Plane 
which has the highest background level at 2600\,MJy\,sr$^{-1}$ at 250\,$\mu$m; Position B 
($l=328.427$\,\degr, $b=+0.04$\,\degr), an unconfused area, also near the centre of the Galactic 
Plane, with a background of 1300\,MJy\,sr$^{-1}$ at 250\,$\mu$m; Position C ($l=328.141$\,\degr, 
$b=-0.89$\,\degr), a confused area near the edge of the Hi-GAL data with a background level of 
450\,MJy\,sr$^{-1}$ at 250\,$\mu$m; and Position D ($l=327.973$\,\degr, $b=-0.99$\,\degr), an unconfused 
area with the lowest background at 65MJy\,sr$^{-1}$ at 250\,$\mu$m.  We define a confused region as one with many nearby
sources. Figures \ref{da_a}--\ref{da_d} shown 
these four regions as they appear in the observations.

Synthetic observations were created of the model IRDCS in each position at \textit{Spitzer} 
8\,$\mu$m and \textit{Herschel} 70, 160, 250, 350 and 500\,$\mu$m. As with the original 
candidates, if the IRDC was seen in emission at the three longest wavelengths it was then classed 
as \textit{Spitzer}-dark and \textit{Herschel}-bright and thus a genuine IRDC. If there was no 
clear emission then it was classed as \textit{Spitzer}-dark and \textit{Herschel}-dark and so 
dismissed. As \textsc{Phaethon} only models the emission of the IRDC, the 8 and 70\,$\mu$m
maps with the added IRDCs appears no different from the original data and so are not necessarily
Spitzer-dark.

\begin{figure*}
\begin{center}
\includegraphics[width=55mm,angle=-90]{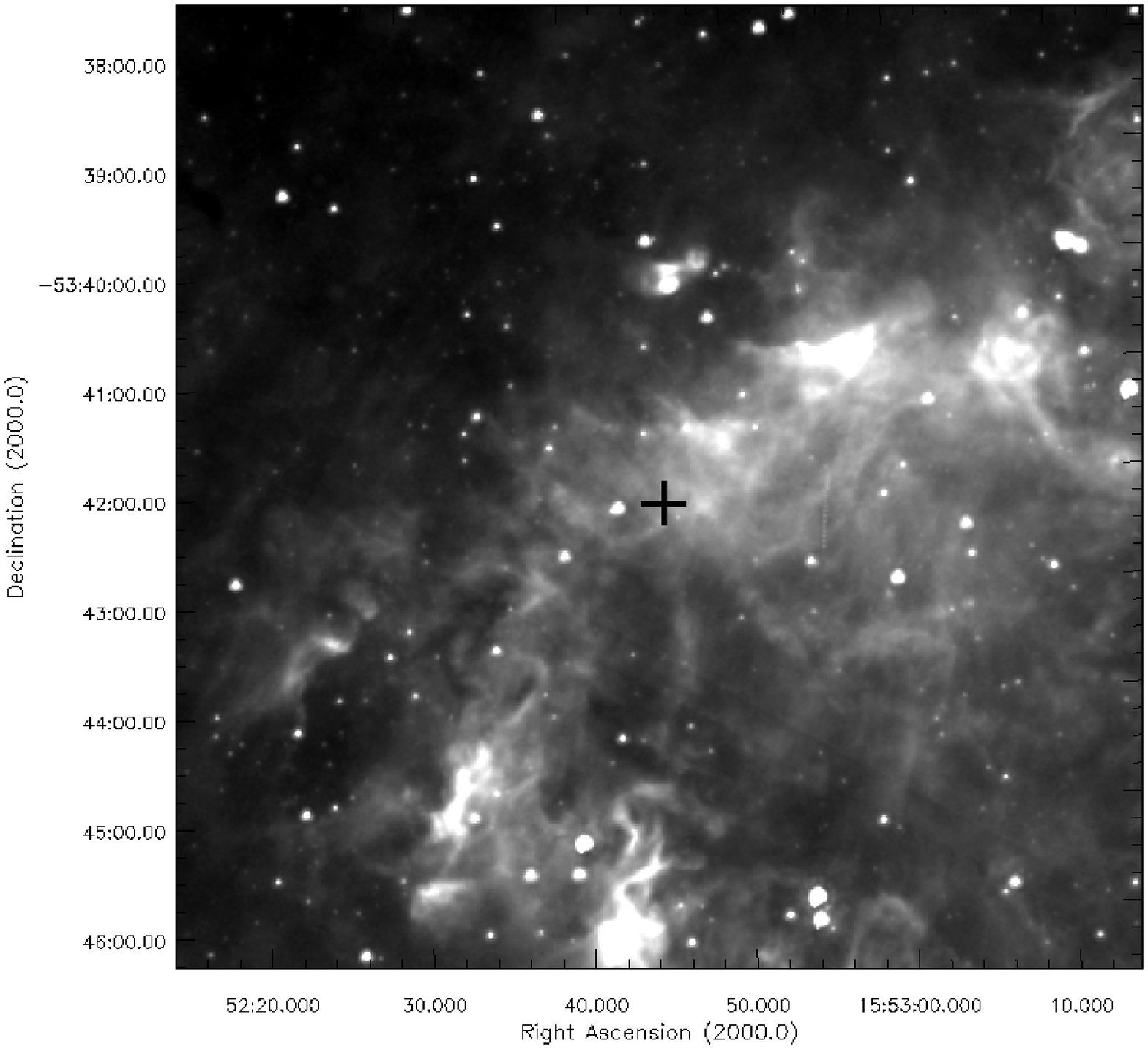}
\includegraphics[width=55mm,angle=-90]{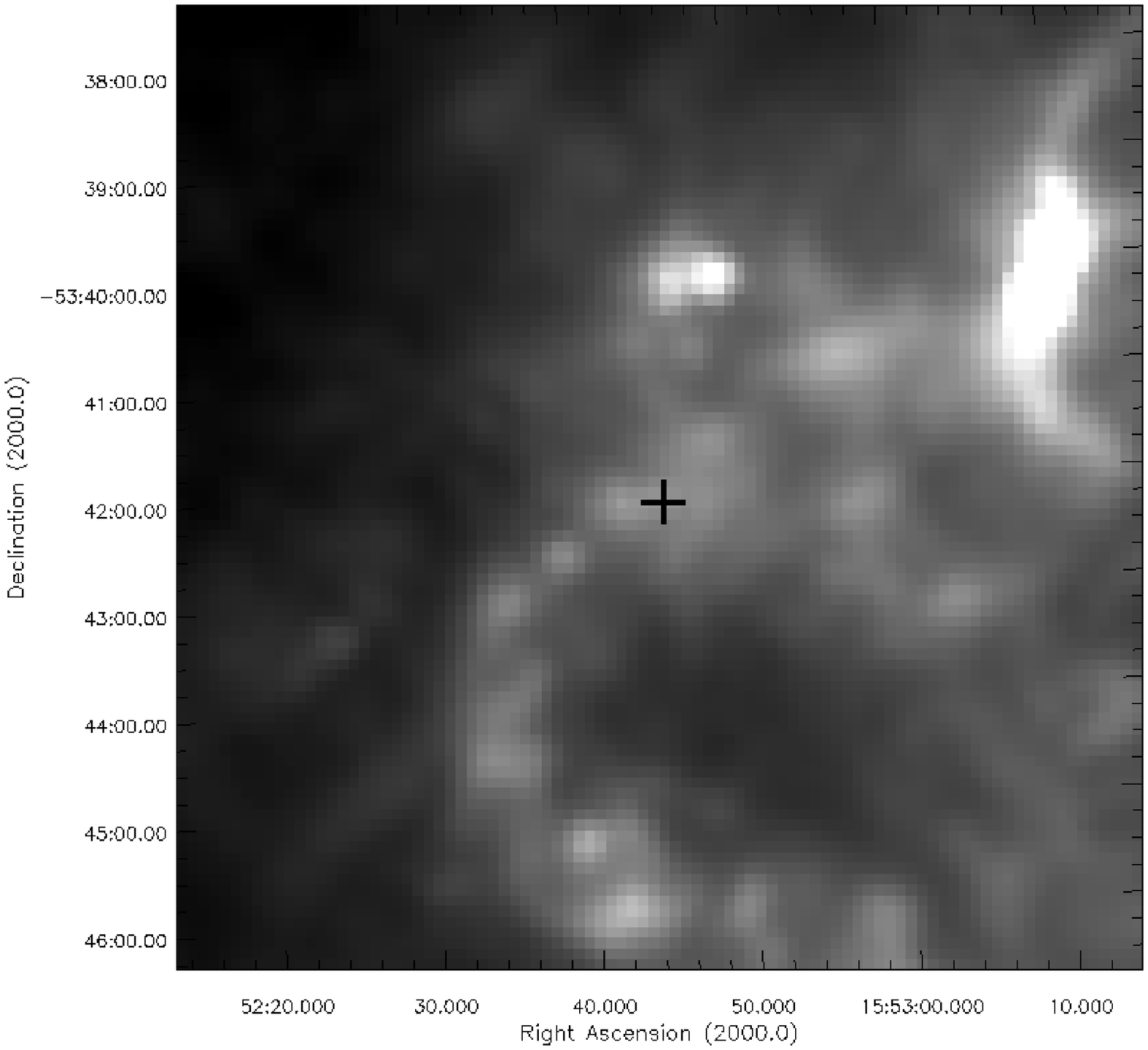}
\end{center}
\caption{A core with radius 0.2\,pc and a peak column density of 2$\times10^{22}$\,cm$^{-2}$
placed in Position A. The IRDC is shown at 8\,$\mu$m (left) and
250\,$\mu$m (right). A cross marks the position of the IRDC.} \label{sl_a}
\end{figure*}

\begin{figure*}
\begin{center}
\includegraphics[width=55mm,angle=-90]{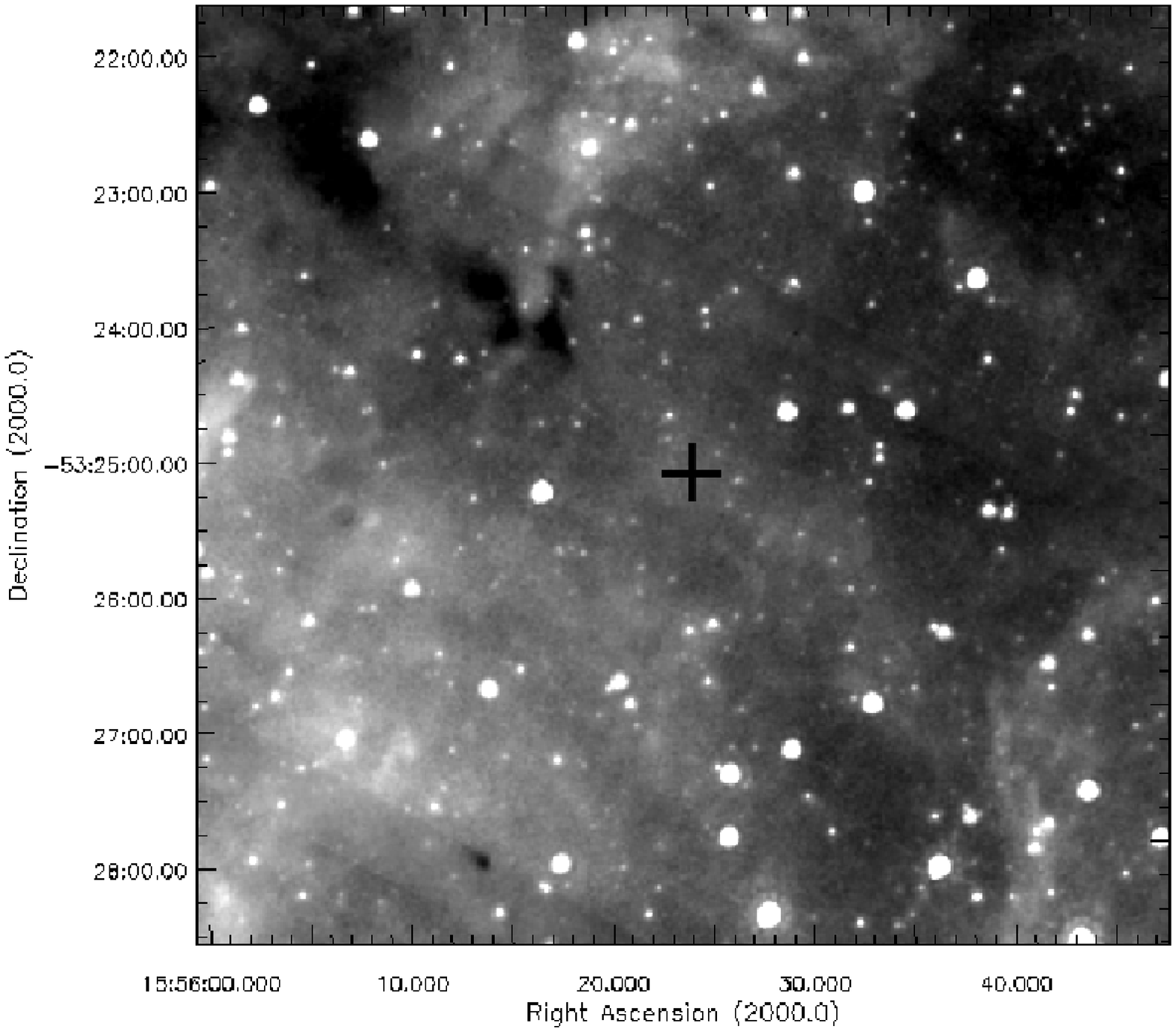}
\includegraphics[width=55mm,angle=-90]{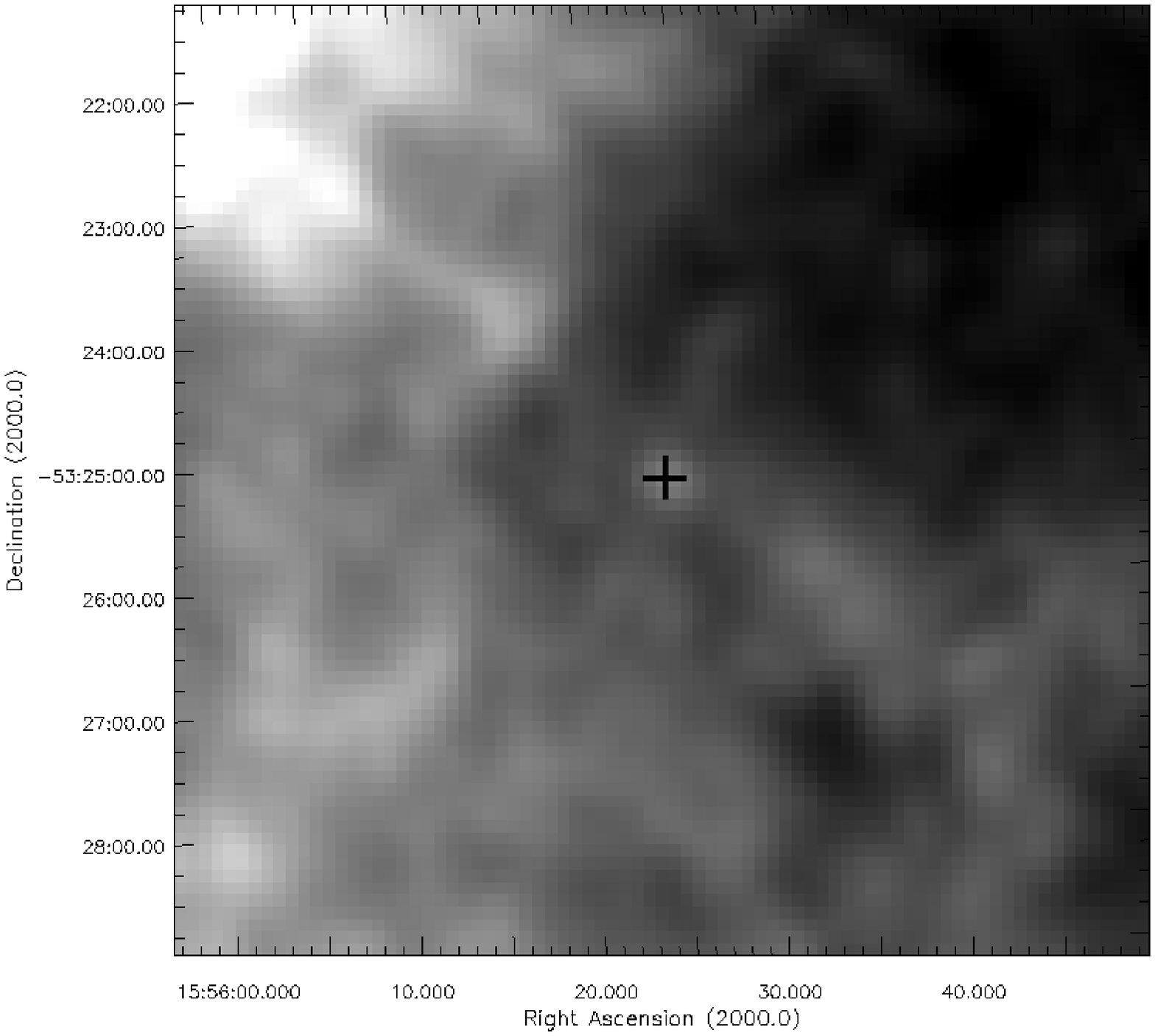}
\end{center}
\caption{A core with radius 0.2\,pc and a peak column density of 2$\times10^{22}$\,cm$^{-2}$
placed in Position B. The IRDC is shown at 8\,$\mu$m (left) and
250\,$\mu$m (right). A cross marks the position of the IRDC.} \label{sl_b}
\end{figure*}

\begin{figure*}
\begin{center}
\includegraphics[width=55mm,angle=-90]{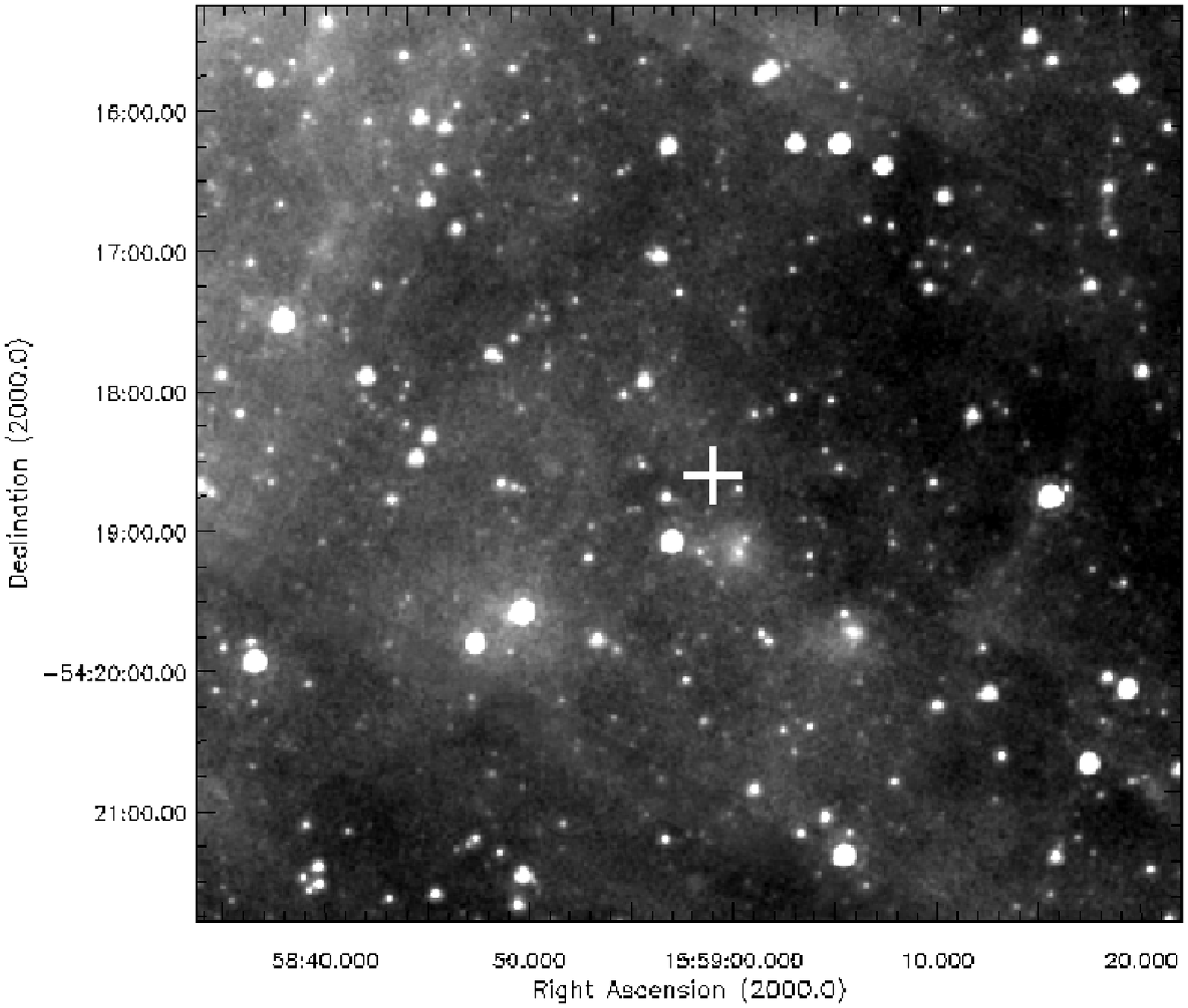}
\includegraphics[width=55mm,angle=-90]{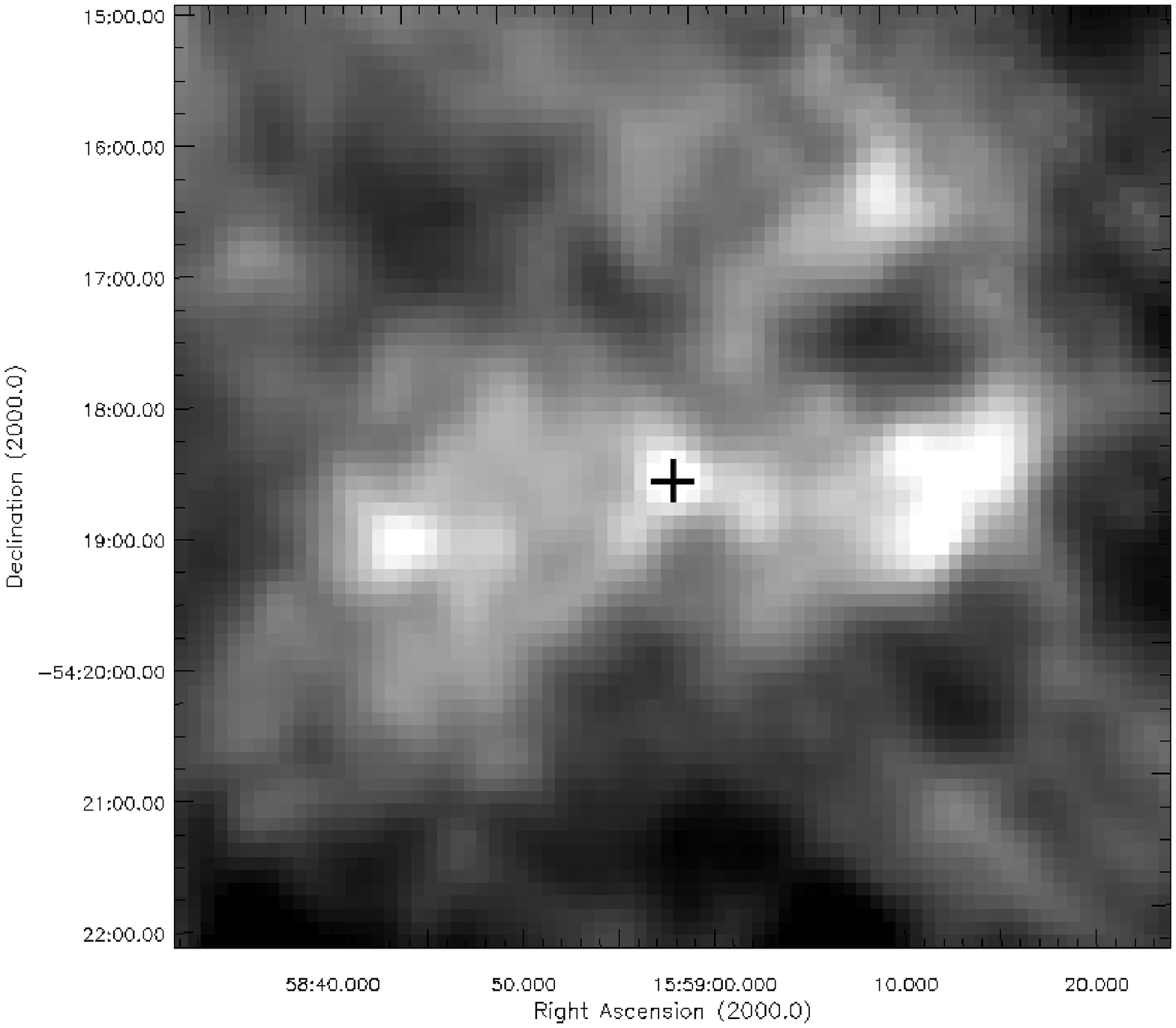}
\end{center}
\caption{A core with radius 0.2\,pc and a peak column density of 2$\times10^{22}$\,cm$^{-2}$
placed in Position C. The IRDC is shown at 8\,$\mu$m (left) and
250\,$\mu$m (right). A cross marks the position of the IRDC.} \label{sl_c}
\end{figure*}

\begin{figure*}
\begin{center}
\includegraphics[width=55mm,angle=-90]{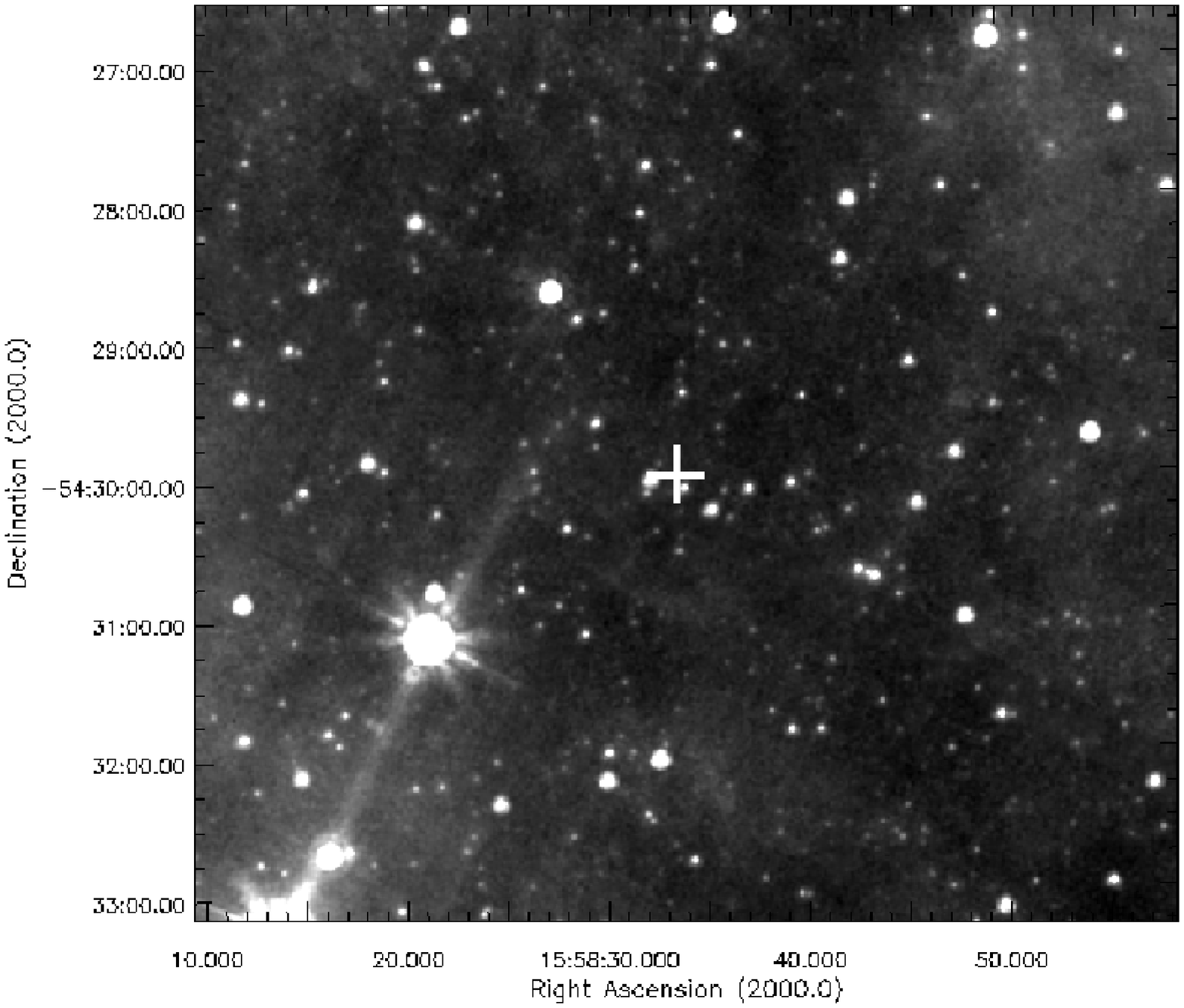}
\includegraphics[width=55mm,angle=-90]{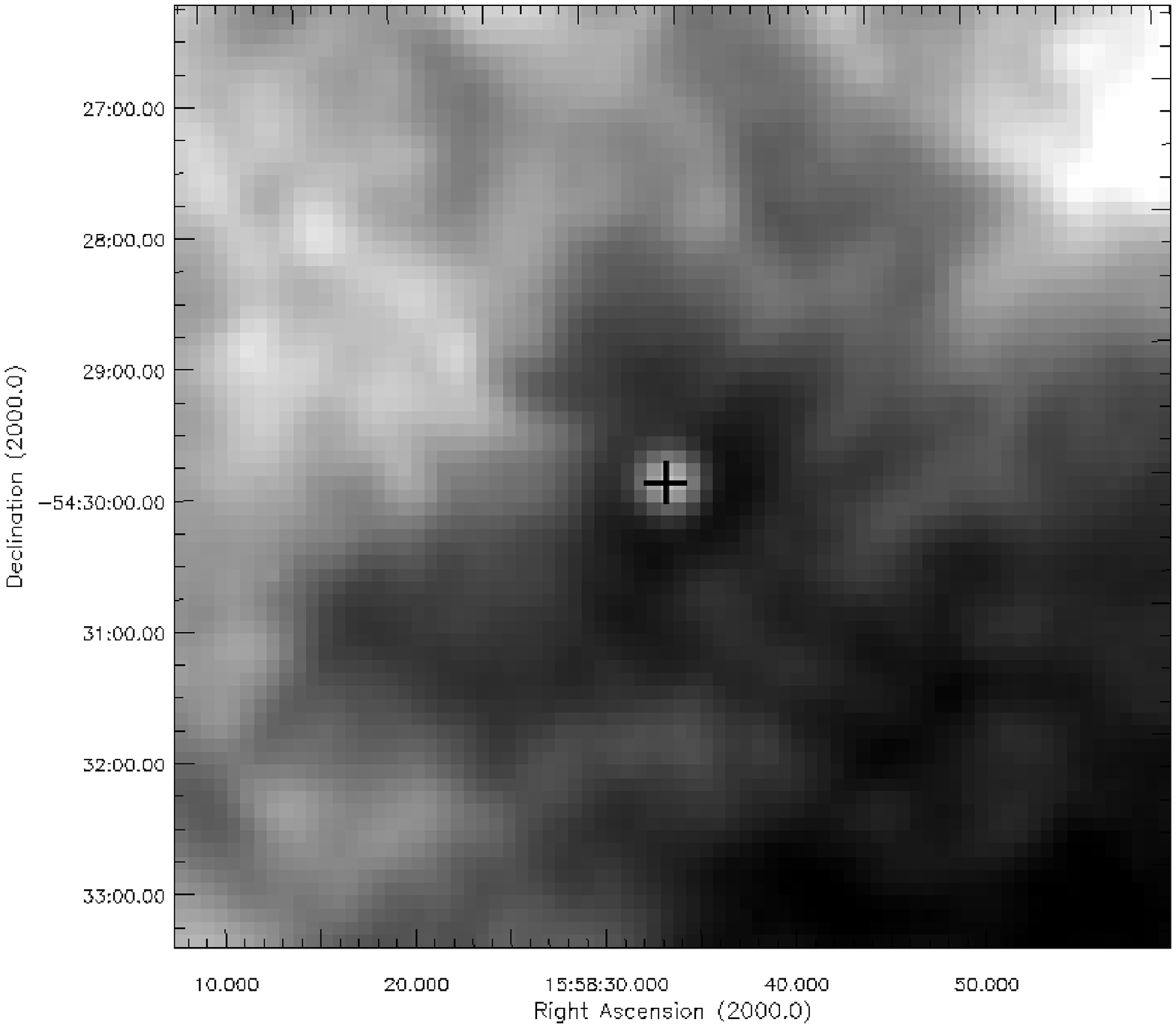}
\end{center}
\caption{A core with radius 0.2\,pc and a peak column density of 2$\times10^{22}$\,cm$^{-2}$
placed in Position D. The IRDC is shown at 8\,$\mu$m (left) and
250\,$\mu$m (right). A cross marks the position of the IRDC.} \label{sl_d}
\end{figure*}

\begin{figure*}
\begin{center}
\includegraphics[width=55mm,angle=-90]{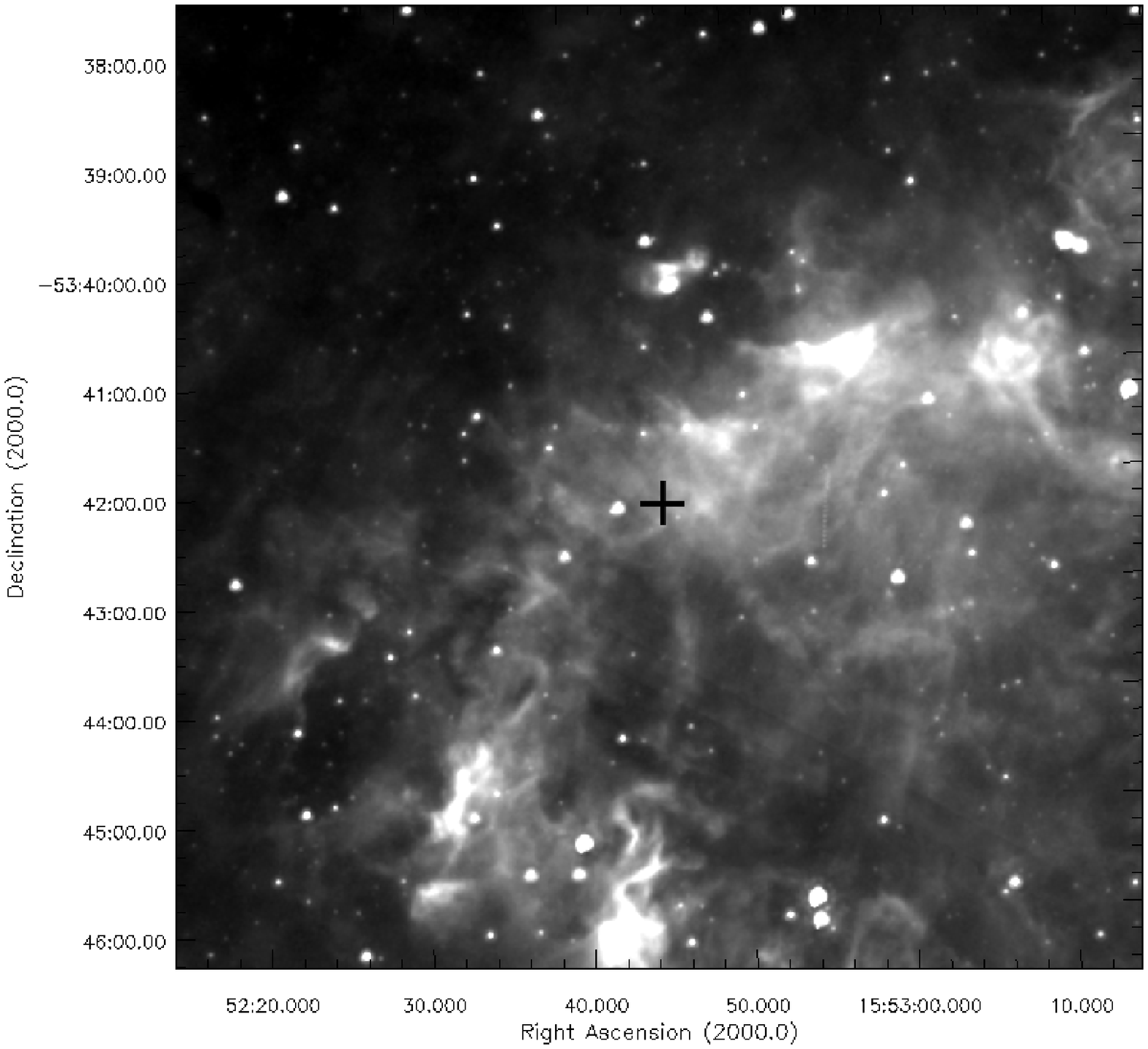}
\includegraphics[width=55mm,angle=-90]{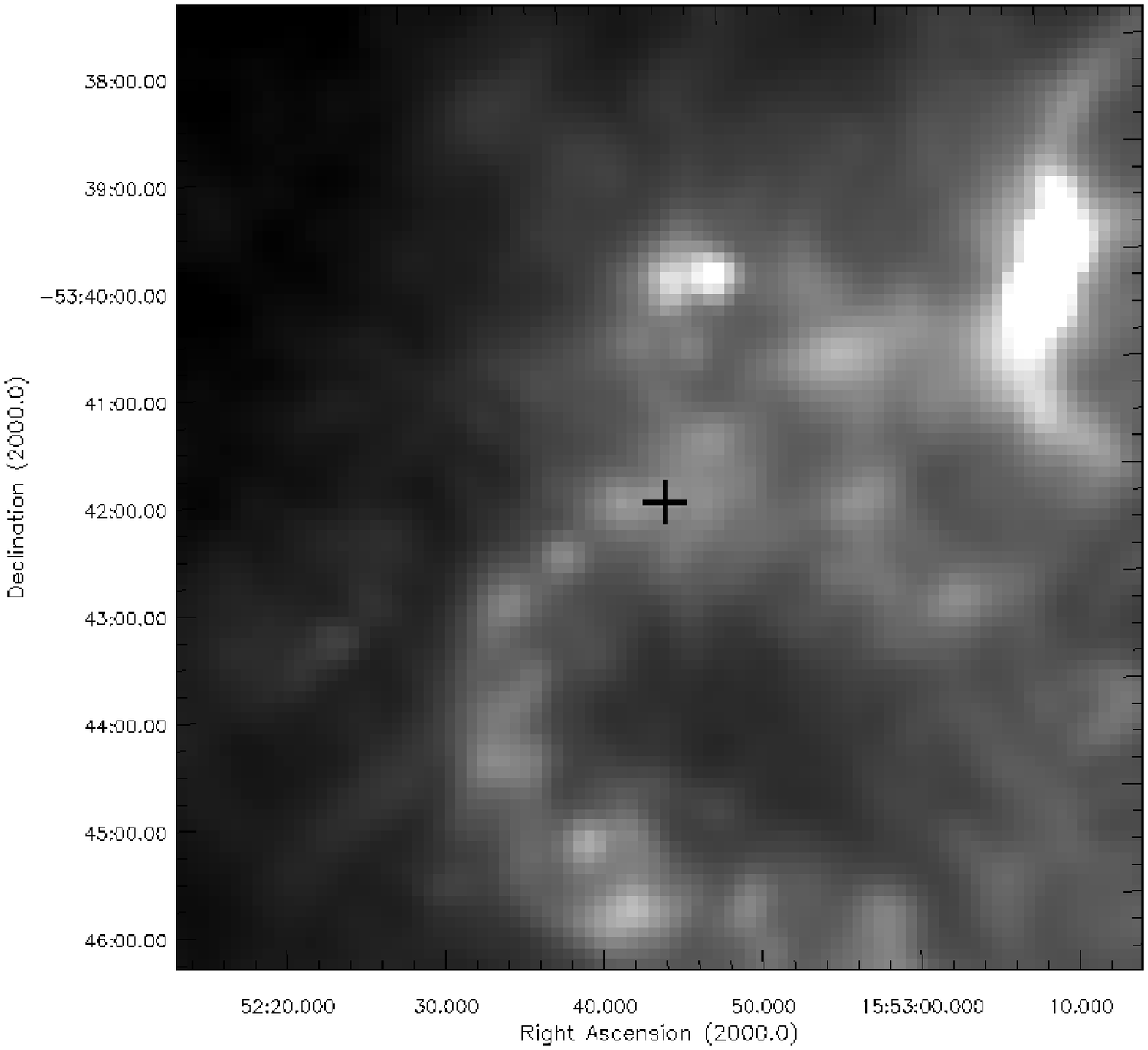}
\end{center}
\caption{A core with radius 0.2\,pc and a peak column density of 4$\times10^{22}$\,cm$^{-2}$
placed in Position A. The IRDC is shown at 8\,$\mu$m (left) and
250\,$\mu$m (right). A cross marks the position of the IRDC.} \label{sh_a}
\end{figure*}

\begin{figure*}
\begin{center}
\includegraphics[width=55mm,angle=-90]{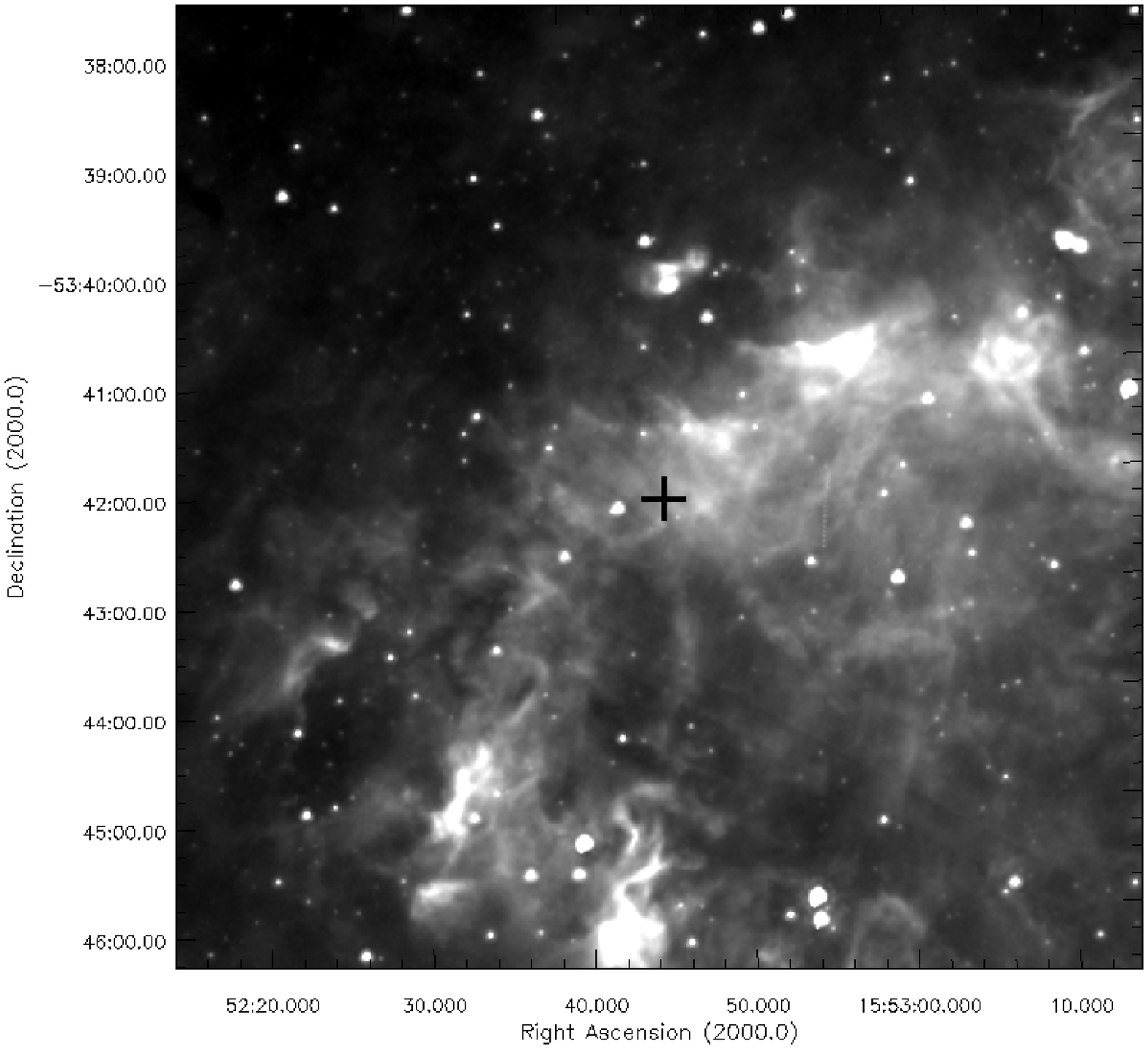}
\includegraphics[width=55mm,angle=-90]{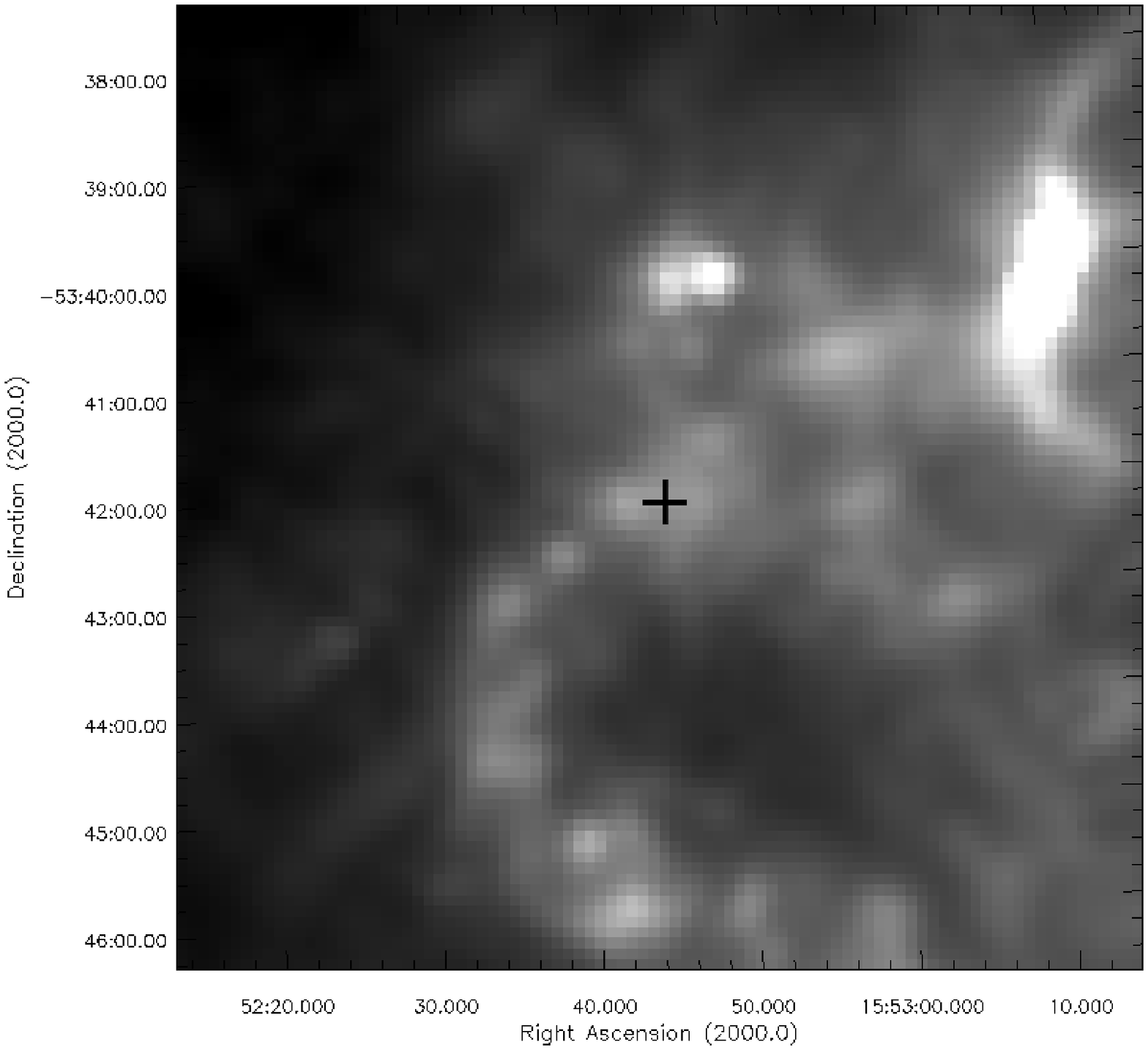}
\end{center}
\caption{A core with radius 0.4\,pc and a peak column density of 2$\times10^{22}$\,cm$^{-2}$
placed in Position A. The IRDC is shown at 8\,$\mu$m (left) and
250\,$\mu$m (right). A cross marks the position of the IRDC.} \label{ml_a}
\end{figure*}

\begin{figure*}
\begin{center}
\includegraphics[width=55mm,angle=-90]{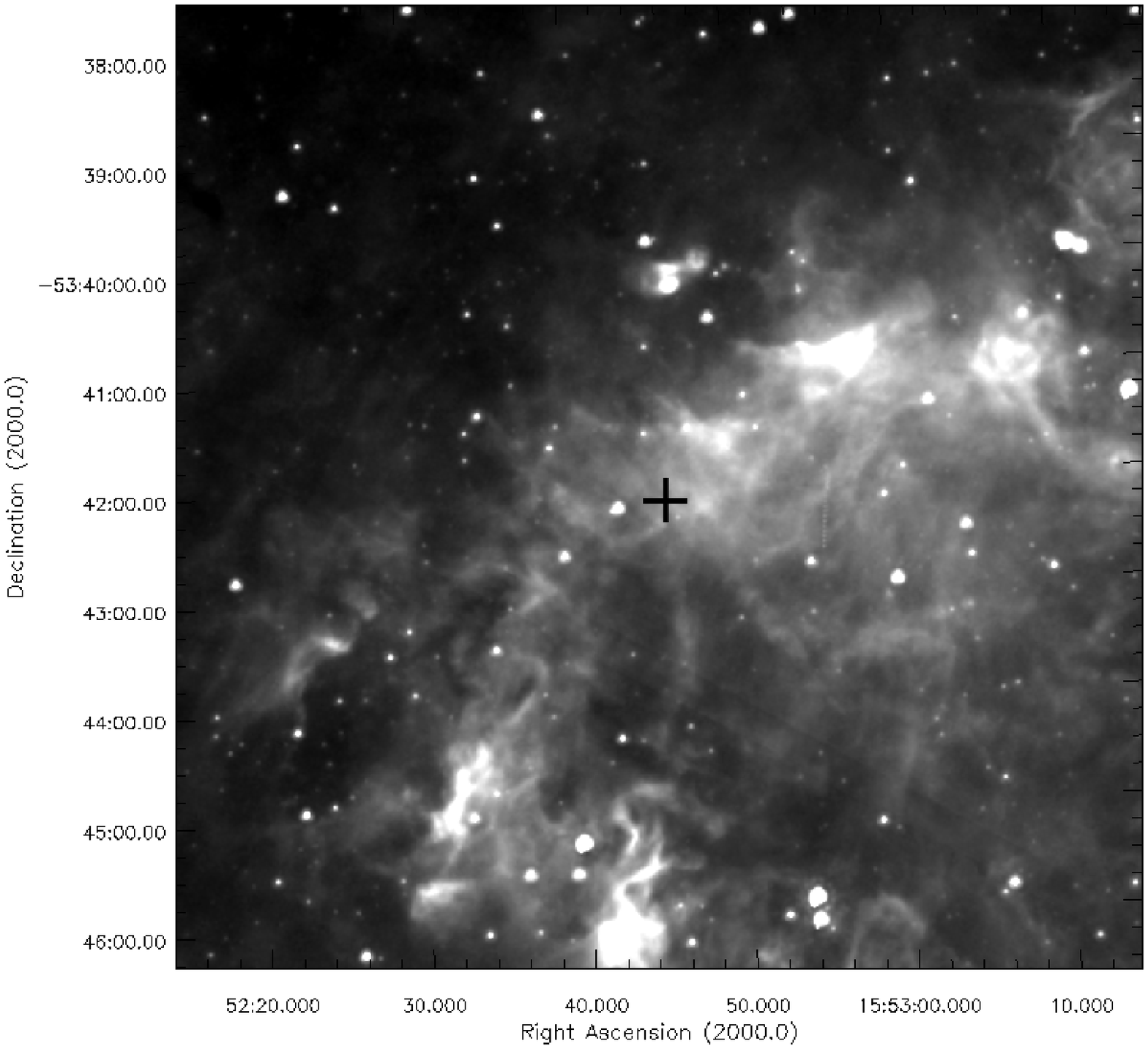}
\includegraphics[width=55mm,angle=-90]{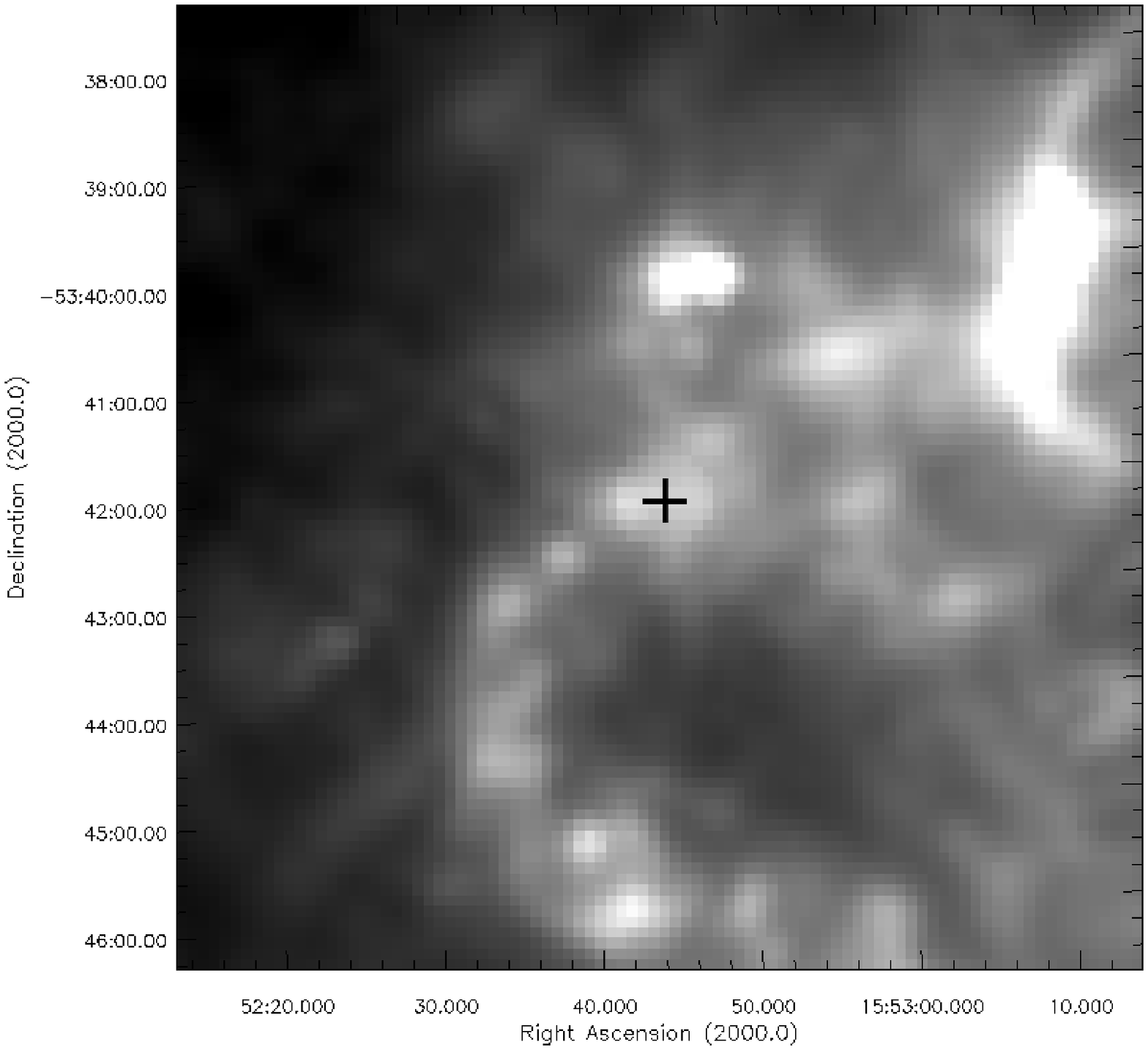}
\end{center}
\caption{A core with radius 0.4\,pc and a peak column density of 4$\times10^{22}$\,cm$^{-2}$
placed in Position A. The IRDC is shown at 8\,$\mu$m (left) and
250\,$\mu$m (right). A cross marks the position of the IRDC.} \label{mh_a}
\end{figure*}

The smallest modelled IRDC (0.2\,pc, 2$\times10^{22}$\,cm$^{-2}$) can clearly be seen in emission
in Positions B, C and D but not in Position A. This can be seen in Figures \ref{sl_a}--\ref{sl_d}. 
A small amount of emission can be seen from the 0.2\,pc IRDC with a peak column density of 
4$\times10^{22}$\,cm$^{-2}$ and the 0.4\,pc, 2$\times10^{22}$\,cm$^{-2}$ IRDC (Figures \ref{sh_a} 
and \ref{ml_a}, respectively) in Position A, although it can not 
be stated with absolute certainty that these objects would have been classed as 
\textit{Herschel}-bright. The smallest IRDC that can clearly be seen in emission at the highest 
background levels is 0.4\,pc in radius with a peak column density of 4$\times10^{22}$\,cm$^{-2}$,
shown in Figure \ref{mh_a}.
We therefore conclude that any IRDC whose major axis is $\leq26$\,\arcsec 
(corresponding to 0.4\,pc at a distance of 3.1\,kpc) with a peak column density less than 
4$\times10^{22}$\,cm$^{-2}$ in an area where the background level is greater than 1300\,MJy\,sr$^{-1}$ at
250\,$\mu$m may not be found to be \textit{Herschel}-bright, regardless of whether it is 
a genuine IRDC or not (although some were). A quantitative analysis of
\textit{Herschel}-bright IRDC detections as a function of their physical parameters will be
addressed in \citet{lenfestey12}.

We then attempted to estimate what percentage of the PF09 catalogue falls below our sensitivity 
limits. We focussed on a 2 degree-square region centred on $l=327$\,\degr, $b=0$\,\degr. This area was 
chosen as it is typical of the $l$ = $300-330$\,\degr\, region and contains many candidate IRDCs. 

The background levels ranged from approximately 390 to 2900\,MJy\,sr$^{-1}$ (equivalent to Positions 
C and A respectively), with an average of 1400\,MJy\,sr$^{-1}$ (equivalent to Position B). The background 
levels were defined using an aperture close to the position of each candidate IRDC. Using axis size and peak opacities from 
PF09 we isolated those candidates 
that fall below our completeness criteria. The column density of each was calculated using:
\begin{equation}
N_{ H_2 } = \tau _{8\,\mu m} \times 3 \left[ \pm 1\right] \times 10^{22} \, \textrm{cm} ^{-2},
\end{equation}
 (PF09), where $N_{ H_2 }$ is the peak column density of the IRDC and $\tau _{8\,\mu m}$
is the peak 8\,$\mu$m opacity taken from PF09. 

This region contains 690 IRDC candidates. 141 of these are \textit{Herschel}-dark with a radius less than 26\,\arcsec, 
a peak column density lower than 4$\times 10^{22}$\,cm$^{-2}$ and a background level above 1300\,MJy\,sr$^{-1}$. 
For these 141 objects we can not state their true status. Therefore, the unknown 
objects comprise $141/690$, or approximately 20\%, of the PF09 IRDC candidates in the $2\times2$\,\degr\, area.

We therefore surmise that, if this pattern is typical of the whole Galactic Plane, then $\sim$20\% are unlikely to be seen 
in emission by \textit{Herschel} regardless of their true status, $\sim$40\% of the PF09 candidate 
IRDCS are \textit{Herschel}-bright, $\sim$40\% are \textit{Herschel}-dark.